\documentclass[journal]{IEEEtran}\IEEEoverridecommandlockouts
\usepackage{graphicx}
\usepackage{amsmath, amsfonts, amssymb, amsthm}
\usepackage{enumerate}
\usepackage{multirow}
\usepackage{multicol}
\usepackage{color}
\usepackage{makecell}
\usepackage{arydshln}
\usepackage{rotating}
\newcommand{\RNum}[1]{\uppercase\expandafter{\romannumeral #1\relax}}
\theoremstyle{definition} \newtheorem{theorem}{Theorem}
\theoremstyle{definition} \newtheorem{corollary}[theorem]{Corollary}
\theoremstyle{definition} \newtheorem{proposition}[theorem]{Proposition}
\theoremstyle{definition} \newtheorem{definition}[theorem]{Definition}
\theoremstyle{definition} \newtheorem{lemma}[theorem]{Lemma}
\theoremstyle{definition} \newtheorem{algorithm}[theorem]{Algorithm}
\theoremstyle{definition} \newtheorem{example}{Example}

\theoremstyle{definition} \newtheorem*{convention}{Conventions}
\theoremstyle{definition} 
\theoremstyle{definition} 
{\end{list}}

\begin{document}

\title{Circular-shift-based Vector Linear Network Coding and Its Application to Array Codes}
\author{\IEEEauthorblockN{Sheng~Jin{$^1$},~Zhe~Zhai{$^1$},~Qifu~Tyler~Sun{$^1$}{$^*$},~Zongpeng~Li{$^2$}} \\
\IEEEauthorblockA{{$^1$}
Department of Communication Engineering, University of Science and Technology Beijing, Beijing, P. R. China \\
{$^2$} Institute for Network Sciences and Cyberspace, Tsinghua University, Beijing, P. R. China \\
}
\thanks{$^*~$Q. T. Sun (Email: qfsun@ustb.edu.cn) is the corresponding author.}}
\maketitle
\sloppy

\begin{abstract}
Circular-shift linear network coding (LNC) is a class of vector LNC with local encoding kernels selected from cyclic permutation matrices, so that it has low coding complexities. %
However, it is insufficient to exactly achieve the capacity of a multicast network, 
so the data units transmitted along the network need to contain redundant symbols, which affects the transmission efficiency. %
In this paper, as a variation of circular-shift LNC, we introduce a new class of vector LNC over arbitrary GF($p$), called \emph{circular-shift-based} vector LNC, which is shown to be able to exactly achieve the capacity of a multicast network. %
The set of local encoding kernels in circular-shift-based vector LNC is nontrivially designed such that it is closed under multiplication by elements in itself. It turns out that the coding complexity of circular-shift-based vector LNC is comparable to and, in some cases, even lower than that of circular-shift LNC. %

The new results in the formulation of circular-shift-based vector LNC further facilitates us to characterize and design Vandermonde circulant maximum distance separable (MDS) array codes, which are built upon the structure of Vandermonde matrices and circular-shift operations. %
We prove that for $r \geq 2$, the largest possible $k$ for an $L$-dimensional $(k+r, k)$ Vandermonde circulant $p$-ary MDS array code is $p^{m_L}-1$, where $L$ is an integer co-prime with $p$, and $m_L$ represents the multiplicative order of $p$ modulo $L$. %
For $r \in \{2, 3\}$, we introduce two new types of $(k+r, k)$ $p$-ary array codes that can achieve the largest $k = p^{m_L}-1$. %
For the special case that $p = 2$, we propose scheduling encoding algorithms for the $2$ new codes, so that the encoding complexity not only asymptotically approaches the optimal $2$ XORs per original data bit, but also slightly outperforms the encoding complexity of other known Vandermonde circulant MDS array codes with largest $k = 2^{m_L} - 1$. 

\end{abstract}

\section{Introduction}
As an extension of conventional scalar linear network coding (LNC), which models data units transmitted along a network over a finite field, vector LNC models data units over a vector space (See, e.g., \cite{ebrahimi2011algebraic}-\cite{chee2020tit}). Compared with scalar LNC, %
vector LNC provides much more candidates of local encoding kernels for coding operations, so that it greatly enhances the flexibility of code design. In the literature, various classical networks have been designed to demonstrate the advantage of vector LNC. For instance, in the M-network constructed in \cite{medard2003coding}, the network capacity can be simply achieved by a $2$-dimensional vector linear network code, but cannot be achieved by scalar LNC over any finite field. %
Specific to (single-source) multicast networks, various instances have been constructed to demonstrate that the size of data units required by vector LNC to achieve the multicast capacity can be smaller than that required by scalar LNC (See, e.g., \cite{sun2016vector}-\cite{han2020tit}). %

Circular-shift LNC is a special class of vector LNC and it was introduced in \cite{Tang_LNC_TIT}, with the aim to reduce the coding complexity. In an $L$-dimensional circular-shift linear network code on a network, the local encoding kernel $\mathbf{K}_{d,e}$ of every adjacent pair $(d,e)$ of edges is selected from $L\times L$ cyclic permutation matrices over GF($2$), so that the encoding operations of circular-shift LNC only contains circular-shifts and bit-wise additions. %
On a binary sequence of length $L$, compared with bit-wise additions, circular-shift operations theoretically incurs negligible computational complexities, so circular-shift LNC has lower encoding and decoding complexities compared with scalar LNC over the extension field GF($2^L$). %
Circular-shift LNC was first formulated in \cite{Tang_LNC_TIT} under the assumption that $L$ is a prime with primitive root $2$, and was extended to be applicable to arbitrary odd dimension $L$ in \cite{Tang_Sun_Circular-shift_LNC_TCOM}. %
In particular, over an arbitrary multicast network, under the assumption that $\frac{m_L}{\phi(L)}2^{m_L}$ is larger than the number of receivers, where $m_L$ and $\phi(L)$ respectively represent the multiplicative order of $2$ modulo $L$ and the Euler's totient function of $L$, an $L$-dimensional circular-shift linear solution at rate $\phi(L)/L$ can be efficiently constructed by the algorithm designed in \cite{Tang_Sun_Circular-shift_LNC_TCOM}. %
Herein, an $L$-dimensional linear solution at rate $\phi(L)/L < 1$ refers to an $L$-dimensional circular-shift linear network code equipped with a $\phi(L) \times L$ matrix $\mathbf{G}$ over GF($2$) subject to: a) every source data unit is a vector of dimension $\phi(L)$ instead of $L$, 
and the outgoing data units of the source are $L$-dimensional vectors generated via multiplying source data units by $\mathbf{G}$; b) every receiver is able to recover all $\phi(L)$-dimensional source data units based on its incoming $L$-dimensional data units. %
For a circular-shift linear solution at rate $\phi(L) / L$, Ref. \cite{Tang_CL19_encoding_decoding_circular_shift_LNC} formulated a general rule to construct the $\phi(L) \times L$ matrix $\mathbf{G}$, and explicitly characterized the decoding process at a receiver, which involves an $L \times \phi(L)$ matrix $\mathbf{H}$ over GF($2$) to transform incoming $L$-dimensional data units into $\phi(L)$-dimensional vectors. %
Random circular-shift LNC was studied in \cite{su2020delay} in the context of wireless broadcast networks. %

In \cite{tang2020circular}, it was proved that even though circular-shift LNC can asymptotically approach the capacity of every multicast network, it is insufficient to exactly achieve the capacity of a multicast network. It was further proved in \cite{tang2022multicast} that even more generalized permutation-based LNC, \emph{i.e.}, vector LNC with local encoding kernels selected from permutation matrices, is not sufficient to exactly achieve the capacity of a multicast network either. %
Consider an $L$-dimensional circular-shift linear solution at rate $\phi(L)/L$, together with the $\phi(L)\times L$ matrix $\mathbf{G}$ used at the source and the $L\times \phi(L)$ matrix $\mathbf{H}$ used at a receiver. Based on the local encoding kernels $\mathbf{K}_{d,e}$ of the considered circular-shift linear network code, a new $\phi(L)$-dimensional vector linear network code with local encoding kernels $\mathbf{G}\mathbf{K}_{d,e}\mathbf{H}$ can be naturally induced. %
If the induced code qualifies as a linear solution, then its rate is $1$ and the capacity of a multicast network can be exactly achieved by a variation of circular-shift LNC. However, justification of whether the induced code is a linear solution is nontrivial and intractable at the first glance. %
In the present work, we first investigate how to formulate a variation of circular-shift LNC that can exactly achieve the capacity of a multicast network, and then apply the new formulation to the design of array codes, which are widely adopted in storage systems to enhance data reliability, such as in Redundant Arrays of Inexpensive Disks (RAID) \cite{patterson1989-raid-introduction}. %
Different from previous study of circular-shift LNC in \cite{Tang_LNC_TIT}-\cite{su2020delay}, which assume the data units to be binary sequences and the code dimension $L$ is an odd prime or an odd integer, in this work, we consider the more general setting that the data units are $p$-ary sequences, and $L$ is an integer co-prime with $p$. %
The main contributions of this paper are summarized as follows:
\begin{itemize}
  \item We introduce a new special class of vector LNC called \emph{circular-shift-based vector LNC} in which the local encoding kernels are selected from the following set $\mathcal{C}$
      \begin{equation}
      \label{eqn: new_local_encoding_kernels_set_brief}
      \mathcal{C} = \mathbf{P}\mathcal{A}\mathbf{Q},
      \end{equation}
      where $\mathcal{A}$ denotes the set of all $L\times L$ circulant matrices over GF($p$), and $\mathbf{P}$, $\mathbf{Q}$ are respectively a $J\times L$ and an $L\times J$ matrix over GF($p$) to be designed. %
      We introduce a general method to construct $\mathbf{P}$ and $\mathbf{Q}$ such that the set $\mathcal{C}$ is closed under multiplication by matrices in it, which is a key to further construct circular-shift-based vector linear solutions. %
      It turns out that the constructed $\mathbf{P}$ and $\mathbf{Q}$ can be selected such that the data units transmitted in a network contain no redundancy compared to circular-shift LNC. Similar to circular-shift LNC, circular-shift-based vector LNC has low coding complexities. In particular, it is demonstrated that in some settings of $J$, $L$ and $\mathbf{P}$, $\mathbf{Q}$, the coding complexity at intermediate nodes of circular-shift-based vector LNC is even lower than that of circular-shift LNC. %
  \item Based on constructed $\mathbf{P}$ and $\mathbf{Q}$ which guarantee the closeness of $\mathcal{C}$ under multiplication by matrices in itself, we further establish an intrinsic connection between circular-shift LNC and circular-shift-based vector LNC. In particular, for every multicast network, an $L$-dimensional circular-shift linear network code with local encoding kernels $\mathbf{K}_{d,e} \in \mathcal{A}$ is a linear solution at rate $J/L$ if and only if the $J$-dimensional circular-shift-based vector linear network code with local encoding kernels $\mathbf{P}\mathbf{K}_{d,e}\mathbf{Q}$ is a linear solution (at rate $1$). It turns out that a circular-shift-based vector linear solution can be efficiently constructed on an arbitrary multicast network. For a circular-shift-based vector linear solution, the decoding matrix at a receiver is explicitly characterized. %
  \item We further investigate circular-shift-based vector LNC on the $(n, k)$-Combination Network, which can be applied to characterize and design array codes. Since $(n, k)$-Combination Network only contains one intermediate node required to perform network coding, it allows us to propose an alternative method to design $\mathbf{P}$ and $\mathbf{Q}$ in the formulation of local encoding kernels in \eqref{eqn: new_local_encoding_kernels_set_brief}. Under the new setting of $\mathbf{P}$ and $\mathbf{Q}$, an intrinsic connection between circular-shift LNC and circular-shift-based vector LNC is also established, and the decoding matrix of a circular-shift-based vector linear solution is explicitly characterized. %
  \item We apply the new results in the formulation of circular-shift-based vector LNC to characterize and design maximum distance separable (MDS) array codes. %
      We focus on the characterization and design of \emph{Vandermonde circulant MDS array codes}, which are built upon the structure of Vandermonde matrices with all encoding kernels belonging to the set \eqref{eqn: new_local_encoding_kernels_set_brief} under certain defined matrices $\mathbf{P}$ and $\mathbf{Q}$. %
      We prove that for $r \geq 2$, the largest possible $k$ for an $L$-dimensional $(k+r, k)$ Vandermonde circulant $p$-ary MDS array code is $p^{m_L}-1$, which was not unveiled in the literature to the best of our knowledge. %
      Equipped with the matrices $\mathbf{P}$ and $\mathbf{Q}$ designed in the formulation of circular-shift-based vector LNC, for $r \in \{2, 3\}$, we introduce two new types of $(k+r, k)$ $p$-ary array codes that can achieve the largest possible $k = p^{m_L}-1$. %
      Compared with other known Vandermonde circulant MDS array codes in the literature, the selection of $\mathbf{P}$ and $\mathbf{Q}$ for forming encoding kernels in the new array codes is much more flexible. %
  \item For the special case that $p = 2$, we propose scheduling algorithms for the encoding process of the $2$ new types of $(k+r, k)$ array codes with largest possible $k = 2^{m_L} - 1$. When $L$ is an odd prime, based on the scheduling algorithms, the encoding complexity of the new array codes not only asymptotically approaches the optimal $2$ XORs per original data bit with increasing $k$ and $L$, but also slightly outperforms the encoding complexity of other known Vandermonde circulant MDS array codes with largest possible $k = 2^{m_L} - 1$. 
\end{itemize}

The remainder of this paper is organized as follows. %
Section \ref{sec:Preliminaries} introduces the network model and reviews the concept of vector LNC and circular-shift LNC. Some useful mathematical notations and lemmas required in this paper are also presented. %
Section \ref{sec:Circular-shift-based vector LNC} formally formulates circular-shift-based vector LNC over general GF($p$) with $p$ a prime. %
Section \ref{sec: Application of the new vector LNC schemes to array codes} first follows the algebraic framework of vector LNC to provide a unified description of array codes, and then applies the new formulation of  circular-shift-based vector LNC to the design of array codes. %
Most technical proofs are placed in Appendix.

\begin{convention}
In this paper, let $p$ be a prime number and $L$ be a positive integer subject to $\gcd(p, L) = 1$. Let $\beta \in \mathrm{GF}(p^{m_L})$ be a primitive $L^{\mathrm{th}}$ root of unity over GF($p$), where $m_L$ represents the multiplicative order of $p$ modulo $L$. %
$\phi(L)$ is the Euler's totient function of $L$, which is the number of integers in $\{1, 2, \ldots, L\}$ that are relatively prime to $L$. %
Every bold symbol represents a vector or a matrix. Let $\mathbf{I}_J$ denote the identity matrix of size $J$, and $\mathbf{0}$, $\mathbf{1}$ respectively denote the all-zero matrix and all-one matrix, %
whose size, if not explicitly explained, can be inferred in the context. $\mathbf{C}_L$ denotes the $L\times L$ cyclic permutation matrix %
$\begin{bmatrix} 
\mathbf{0} & \mathbf{I}_{L-1} \\ 
1 & \mathbf{0} \end{bmatrix}$. %
$[\mathbf{A}_e]_{e \in E}$ refers to the block matrix obtained by column-wise juxtaposition of matrices $\mathbf{A}_e$ with $e$ chosen from a set $E$, %
where $\mathbf{A}_e$ may degenerate to vectors or scalars. Moreover, $[\mathbf{A}_{m, n}]_{1 \leq m \leq M, 1 \leq n \leq N}$ refers to the $M \times N$ block matrix, in which every block $\mathbf{A}_{m, n}$ is the block entry %
with row and column indexed by $m$ and $n$ respectively. %
The Kronecker product is denoted by $\otimes$. For an arbitrary $l_1\times l_2$ matrix $\mathbf{A}$, let $\mathbf{A}_{l\otimes}$ denote the $l_1l\times l_2l$ matrix $\mathbf{I}_l\otimes \mathbf{A}$.
\end{convention}

\section{Preliminaries}
\label{sec:Preliminaries}
\subsection{Network model}
In this paper, we formulate circular-shift-based vector LNC over a multicast network, which is modeled as a finite directed acyclic multigraph, with a unique source node $s$ and a set $T$ of receivers. Similar formulation can be readily extended to multi-source multicast networks, which will not be considered in this paper for conciseness. %
For a node $v$ in a network, the set of its incoming and outgoing edges are respectively denoted by $\mathrm{In}(v)$ and $\mathrm{Out}(v)$. If there is a node $v$ with $d \in \mathrm{In}(v)$ and
$e \in \mathrm{Out}(v)$, the pair ($d, e$) is called an adjacent pair. Every edge has a unit 
capacity to transmit a data unit, which is an $L$-dimensional row vector $\mathbf{m}_e$ of $p$-ary data symbols. Write $|\mathrm{Out}(s)| = \omega$. 
The source node $s$ generates $\omega$ independent data units and transmits them along the network. The process that every intermediate node recombines incoming data units to generate outgoing data units, is called \emph{recoding}. For every receiver $t \in T$, it aims to recover 
$\omega$ data units generated by the source $s$. Without loss of generality, assume that there is not any edge leading from $s$ to $t$, 
and $|\mathrm{Out}(s)| = |\mathrm{In}(t)| = \omega$.

\subsection{Vector LNC and circular-shift LNC}
In conventional scalar LNC, the data unit $\mathbf{m}_e$ transmitted along every edge is regarded as an element in the finite field GF($p^L$). Vector LNC (See, e.g., \cite{ebrahimi2011algebraic}) is an extension of conventional scalar LNC and it models every data unit $\mathbf{m}_e$ as an $L$-dimensional vector over GF($p$). %
Specifically, an $L$-dimensional vector linear code over GF($p$) is an assignment of a \emph{local encoding kernel} $\mathbf{K}_{d, e}$, which is an $L \times L$ matrix over GF($p$), 
to every pair ($d, e$) of edges such that $\mathbf{K}_{d, e}$ is the zero matrix $\mathbf{0}$ when ($d, e$) is not an adjacent pair. For a non-source node $v$ and $e \in \mathrm{Out}(v)$, 
the data unit can be expressed as $\mathbf{m}_e = \sum\nolimits_{d \in \mathrm{In}(v)}\mathbf{m}_d \mathbf{K}_{d, e}$. Every vector linear code uniquely determines a global encoding kernel $\mathbf{F}_e$, which is an 
$\omega L \times L$ matrix over GF($p$). For every edge $e$ emanating from source $s$, the block matrix $[\mathbf{F}_e]_{e \in \mathrm{Out(s)}} = \mathbf{I}_{\omega L}$, and for every outgoing edge $e$ from 
non-source node $v$, $\mathbf{F}_e = \sum\nolimits_{d \in \mathrm{In(v)}}\mathbf{F}_d\mathbf{K}_{d, e}$. Correspondingly, the data unit $\mathbf{m}_e$ can also be expressed as $\mathbf{m}_e = [\mathbf{m}_d]_{d \in \mathrm{Out(s)}}\mathbf{F}_e$. %
A vector linear code is qualified to be a \emph{vector linear solution} if for every receiver $t \in T$, the $\omega L\times \omega L$ matrix $[\mathbf{F}_e]_{e \in \mathrm{In(t)}}$ over GF($p$) has full rank $\omega L$.
The advantages of vector LNC over scalar LNC been extensively studied in the literature (See, e.g., \cite{ebrahimi2011algebraic}-\cite{chee2020tit},\cite{Tang_LNC_TIT}). 

In \cite{Tang_LNC_TIT}-\cite{su2020delay}, a special class of vector LNC, called circular-shift LNC, was formulated over GF($2$). Herein, we extend the formulation over GF($p$). An $L$-dimensional circular-shift linear network code ($\mathbf{K}_{d,e}$) over GF($p$) is an $L$-dimensional vector linear code with local encoding kernels selected from 
\begin{equation}
\label{eqn: circular-shift_kernel}
\mathcal{A} = \left\{\sum\nolimits_{j=0}^{L-1}a_j\mathbf{C}_L^j: a_j \in \mbox{GF($p$)} \right\},
\end{equation} %
which consists of all $L\times L$ circulant matrices over GF($p$). %
The reason to investigate circular-shift LNC is low computational complexity to perform the coding operations, which only consist of circular-shifts among $L$ $p$-ary symbols and symbol-wise addition. However, it has been proved in \cite{Tang_LNC_TIT} that circular-shift LNC is insufficient to exactly achieve the capacity of a multicast network, 
that is, there exists a multicast network such that it does not have an $L$-dimensional circular-shift linear network code over GF($p$) for any $L$ and prime $p$ if the local encoding kernels are selected from \eqref{eqn: circular-shift_kernel}. %
It turns out that a circular-shift linear solution at rate smaller than $1$ is inevitable, that is, circular-shift LNC is insufficient to achieve the exact multicast capacity. %
Specifically, for an $L$-dimensional circular-shift linear solution over GF($p$) at rate $L'/L$ with $L' < L$, the data units transmitted on every edge in the network are still $L$-dimensional row vector over GF($p$). The difference is that the source $s$ generates $\omega$ source data units $\mathbf{m}'_1, \ldots, \mathbf{m}'_{\omega}$, each of which is an $L'$-dimensional row vector over GF($p$). %
Based on these $\omega$ source data units and an $\omega L' \times \omega L$ matrix $\mathbf{G}_s$ over GF($p$), called the source encoding matrix, the  $\omega$ data units $\mathbf{m}_e$ transmitted on outgoing edges $e$ of $s$ can be computed as 
\begin{equation}
\label{eqn: circular-shift_LNC_encoding_matrix}
[\mathbf{m}_e]_{e \in \mathrm{Out(s)}} = [\mathbf{m}'_i]_{1 \leq i \leq \omega}\mathbf{G}_s,
\end{equation}
When $L$ is an odd integer with $\frac{m_L}{\phi(L)}2^{m_L} > |T|$, an $L$-dimensional circular-shift linear solution over GF($2$) at rate $\phi(L)/L$ can be efficiently constructed based on the algorithm proposed in \cite{Tang_Sun_Circular-shift_LNC_TCOM}. %

Even though circular-shift LNC allows a lower computational complexity for the data transmission with only circular-shifts and symbol-wise addition, so as to guarantee the decodability at receivers, 
it is inevitable to carry redundancy symbols during the transmission, which affects the transmission efficiency. Circular-shift-based vector LNC to be formulated in this paper is an extension of circular-shift LNC by eliminating transmitted redundant symbols, so that a vector linear solution at rate $1$ and with low computational complexity can be constructed. %
In addition, a general framework to construct a source encoding matrix and a corresponding decoding matrix at every receiver for the circular-shift linear network code over GF($2$) was proposed in \cite{Tang_CL19_encoding_decoding_circular_shift_LNC}. %

\subsection{Mathematical notations and useful lemmas}
\label{subsec:Mathematical notations and useful lemmas}
In the remainder of this paper, adopt the following notation and assumptions. Assume that $L$ an integer co-prime with $p$. Denote by $m_L$ the multiplicative order of $p$ modulo $L$, and $L_{\mathrm{GF}(p)}$ denote the element in GF($p$) represented by $L$ modulo $p$. %
Let $\beta \in \mathrm{GF}(p^{m_L})$ be a primitive $L^{\mathrm{th}}$ root of unity over GF($p$). Denote by $\mathcal{J}$ a subset of $\{0, 1, 2, \ldots, L-1\}$ that is closed under multiplication by $p$ modulo $L$. 
Consequently, $\prod_{j\in \mathcal{J}} (x-\beta^j)$ is a polynomial over GF($p$) that factors $x^L - 1$. Let $J$ be the cardinality of $\mathcal{J}$.

In the course of formulation circular-shift LNC in \cite{Tang_LNC_TIT}-\cite{Tang_CL19_encoding_decoding_circular_shift_LNC}, an important mathematical tool is to decompose the cyclic permutation matrix $\mathbf{C}_L$ in the following way (See, e.g., \cite{tang2020circular})
\begin{equation}
\label{eqn:diagonalization}
\mathbf{C}_L^j=L_{\mathrm{GF}(p)}^{-1}\mathbf{V}_L\mathbf{\Lambda}^j\widetilde{\mathbf{V}}_L, ~~\forall j \geq 0.
\end{equation}
In \eqref{eqn:diagonalization}, $\mathbf{V}_L$ and $\widetilde{\mathbf{V}}_L$ are the $L \times L$ Vandermonde matrix over GF($p^{m_L}$) respectively generated by $1, \beta, \beta^2, \ldots, \beta^{L-1}$ and $1, \beta^{-1}, \beta^{-2}, \ldots, \beta^{-(L-1)}$, %
that is, 
\begin{equation}
\label{eqn: VL}
\mathbf{V}_L = \begin{bmatrix}
1 & 1 & 1 & \ldots & 1 \\
1 & \beta & \beta^2 & \ldots & \beta^{L-1} \\
\vdots & \vdots & \vdots & \ldots & \vdots \\
1 & \beta^{L-1} & \beta^{2(L-1)} & \ldots & \beta^{(L-1)(L-1)}
\end{bmatrix}, 
\end{equation}
\begin{equation}
\label{eqn: widetilde_VL}
\widetilde{\mathbf{V}}_L = \begin{bmatrix}
1 & 1 & 1 & \ldots & 1 \\
1 & \beta^{-1} & \beta^{-2} & \ldots & \beta^{-(L-1)} \\
\vdots & \vdots & \vdots & \ldots & \vdots \\
1 & \beta^{-(L-1)} & \beta^{-(L-1)2} & \ldots & \beta^{-(L-1)(L-1)}
\end{bmatrix}, 
\end{equation}
and $\mathbf{\Lambda}$ is the $L \times L$ diagonal matrix with diagonal entries equal to $1, \beta, \ldots, \beta^{L-1}$. %

\begin{lemma}
\label{lemma: properties_on_VL_and_tilde_VL}
For the Vandermonde matrix $\mathbf{V}_L$ and $\widetilde{\mathbf{V}}_L$ mentioned above, we have the following properties:
\begin{equation}
\label{eqn:VL_VL_tilde}
\mathbf{V}_L\widetilde{\mathbf{V}}_L = \widetilde{\mathbf{V}}_L\mathbf{V}_L = L_{\mathrm{GF}(p)}\mathbf{I}_L,
\end{equation}
\begin{equation}
\label{eqn:VL_tilde^2}
\mathbf{V}_L^2 = \widetilde{\mathbf{V}}_L^{2} = L_{\mathrm{GF}(p)}\begin{bmatrix}
1&0 & \ldots & 0\\
0&0 & \ldots & 1\\
\vdots & \vdots & \begin{sideways} $\ddots$ \end{sideways} &\vdots  \\
0&1&\ldots  &0
\end{bmatrix}.
\end{equation}
\end{lemma}
\begin{IEEEproof}
Based on the definition of $\mathbf{V}_L$ and $\widetilde{\mathbf{V}}_L$, it can be readily observed that %
\begin{align}
\label{eqn:VL_VL_tilde^2_in_sum}
\mathbf{V}_L\widetilde{\mathbf{V}}_L = [\sum\nolimits_{l=0}^{L-1}\beta^{(i-j)l}]_{0\leq i, j \leq L-1}, \\
\label{eqn:VL_tilde^2_in_sum}
\mathbf{V}_L^2 = [\sum\nolimits_{l=0}^{L-1}\beta^{(i+j)l}]_{0\leq i, j \leq L-1}.
\end{align}

First we prove \eqref{eqn:VL_VL_tilde} based on \eqref{eqn:VL_VL_tilde^2_in_sum}. When $i=j$, $\sum\nolimits_{l=0}^{L-1}\beta^{(i-j)l} = \sum\nolimits_{l=0}^{L-1}1$, which is equal to $L_{\mathrm{GF}(p)}$. %
Since $\beta$ is a primitive $L^{th}$ root of unity over GF($p$), $\beta^0, \beta^1, \ldots, \beta^{L-1}$ constitute the $L$ distinct roots of $x^L-1 = (x-1)(x^{L-1}+x^{L-2}+\ldots+1)$. Consequently, when $i \neq j$, $\sum\nolimits_{l=0}^{L-1}\beta^{(i-j)l} = 0$. %
As a result, $\mathbf{V}_L\widetilde{\mathbf{V}}_L = L_{\mathrm{GF}(p)}\mathbf{I}_L$. Since $\beta^{-1}$ is also an element over GF($p^{m_L}$) with multiplicative order $L$, by an exactly same approach, we can deduce $\widetilde{\mathbf{V}}_L\mathbf{V}_L = L_{\mathrm{GF}(p)}\mathbf{I}_L$. %

We next prove \eqref{eqn:VL_tilde^2} based on \eqref{eqn:VL_tilde^2_in_sum}. When $i+j \equiv 0 \bmod L$, $\sum\nolimits_{l=0}^{L-1}\beta^{(i+j)l} = \sum\nolimits_{l=0}^{L-1}1$, which is equal to $L_{\mathrm{GF}(p)}$. %
Analogous to the proof of \eqref{eqn:VL_VL_tilde}, when $i+j \neq 0$, $\sum\nolimits_{l=0}^{L-1}\beta^{(i+j)l} = \sum\nolimits_{l=0}^{L-1}\beta^{-(i+j)l} = 0$. It turns out that $\mathbf{V}_L^2 = \widetilde{\mathbf{V}}_L^2 = L_{\mathrm{GF}(p)}\begin{bmatrix}\begin{smallmatrix}
1&0 & \ldots & 0\\
0&0 & \ldots & 1\\
\vdots & \vdots & \begin{sideways} $\ddots$ \end{sideways} &\vdots  \\
0&1&\ldots  &0
\end{smallmatrix}\end{bmatrix}$.
\end{IEEEproof}

\section{Circular-shift-based vector LNC}
\label{sec:Circular-shift-based vector LNC}
Analogous to the formulation of circular-shift LNC (See, e.g., \cite{Tang_LNC_TIT}-\cite{Tang_CL19_encoding_decoding_circular_shift_LNC}), in this section, we formulate a general design framework for circular-shift-based vector LNC with local encoding kernels selected from a set in the form of %
\begin{equation}
\label{eqn:circular-shift-based_kernel}
\mathcal{C} = \left\{\mathbf{P}(\sum\nolimits_{j=0}^{L-1}a_j\mathbf{C}_L^j)\mathbf{Q}: a_j \in \mathrm{GF}(p) \right\},
\end{equation}
where $\mathbf{P}$ and $\mathbf{Q}$ are $J \times L$ and $L \times J$ matrices over GF($p$) respectively. %
Recall that circular-shift LNC \cite{Tang_LNC_TIT}-\cite{Tang_CL19_encoding_decoding_circular_shift_LNC}, whose local encoding kernels are selected from $\{\sum\nolimits_{j=0}^{L-1}a_j\mathbf{C}_L^j: a_j \in \mathrm{GF}(p)\}$ in \eqref{eqn: circular-shift_kernel}, is not sufficient to exactly achieve the multicast capacity, so a circular-shift linear solution at rate smaller than $1$ has to be constructed. %
As to be clear in this section, the newly constructed circular-shift-based vector LNC can exactly achieve the multicast capacity, that is, for an arbitrary multicast network, there is a proper selection of $\mathbf{P}$, $\mathbf{Q}$ and $L$ such that a vector linear solution with local encoding kernels selected from $\mathcal{C}$ can be efficiently constructed. %

We refer to a vector linear network code with local encoding kernels selected from $\mathcal{C}$ in \eqref{eqn:circular-shift-based_kernel} as a circular-shift-based linear network code because in the recoding process, it not only involves the circular-shift operations, but also the linear operations induced by $\mathbf{P}$ and $\mathbf{Q}$. In the remaining part of this section, we shall first introduce a proper design of matrices $\mathbf{P}$ and $\mathbf{Q}$ such that $\mathcal{C}$ remains closed under multiplication by matrices in $\mathcal{C}$, and every nonzero matrix in $\mathcal{C}$ is invertible. %
Then, we shall discuss how to construct a circular-shift-based vector linear solution that exactly achieves the multicast capacity. The corresponding decoding matrix at every receiver will also be explicitly characterized. In addition, we employ circular-shift-based vector LNC to construct linear solutions on the $(n, k)$-Combination Network.

Hereafter in this paper, for a polynomial $f(x)$ over $\mathrm{GF}(p)$, let $f(\mathbf{C}_L)$ denote the $L\times L$ circulant matrix obtained by evaluation of $f(x)$ under the setting $x = \mathbf{C}_L$. %


\subsection{Explicit construction of $\mathbf{P}$ and $\mathbf{Q}$}
\label{subsec:GH}
In this subsection, we shall introduce a general method to construct a $J\times L$ matrix $\mathbf{P}$ and an $L \times J$ matrix $\mathbf{Q}$ such that the set $\mathcal{C}$ defined in \eqref{eqn:circular-shift-based_kernel}
is closed under multiplication by matrices in it, which is a key to further construct circular-shift-based vector linear solutions.

Recall that $\mathcal{J}$ denotes a subset of $\{0, 1, 2, \ldots, L-1\}$ that is closed under multiplication by $p$ modulo $L$, and $J$ denotes the cardinality of $\mathcal{J}$. %
Consider two $J\times L$ matrices $\mathbf{U} = [\mathbf{u}_j]_{0\leq j \leq L-1}$ and $\mathbf{U}' = [\mathbf{u}_j']_{0\leq j \leq L-1}$ over GF($p^{m_L}$) subject to the following
\begin{align}
\label{U'_def}
\mathrm{rank}([\mathbf{u}_j]_{j\in {\cal J}}) = J, &~ [\mathbf{u}'_j]_{j\in \cal J}^{\mathrm{T}} = L_{\mathrm{GF}(p)}^{-1}[\mathbf{u}_j]_{j\in {\cal J}}^{-1} \\
\label{j notin J}
[\mathbf{u}_j]_{j\notin \cal J} = &\mathbf{0}~\mathrm{or}~[\mathbf{u}_j']_{j \notin \cal J} = \mathbf{0}.
\end{align}
The desired $J\times L$ matrix $\mathbf{G}$ and the $L \times J$ matrices $\mathbf{H}$ are defined by
\begin{equation}
\label{eqn:GH_def}
\mathbf{G} = \mathbf{U}\widetilde{\mathbf{V}}_L, \mathbf{H}=\mathbf{V}_L\mathbf{U}'^{\mathrm{T}},
\end{equation}
where $\mathbf{V}$ and $\widetilde{\mathbf{V}}$ are respectively defined in \eqref{eqn: VL} and \eqref{eqn: widetilde_VL}. %
By \eqref{U'_def}, $\mathbf{G}\mathbf{H} = \mathbf{I}_J$. Notice that the matrices $\mathbf{G}$ and $\mathbf{H}$ are not necessarily over GF($p$), but we can show the following.

\begin{proposition}
\label{prop: G H GF(p)}
If $\mathbf{G}$ is over GF($p$) and $[\mathbf{u}_j']_{j \notin \cal J} = \mathbf{0}$, then the counterpart $L\times J$ matrix $\mathbf{H}$ is over GF($p$).
Conversely, if $\mathbf{H}$ is over GF($p$) and $[\mathbf{u}_j]_{j \notin \cal J} = \mathbf{0}$, then the counterpart $J\times L$ matrix $\mathbf{G}$ is over GF($p$).
\begin{IEEEproof}
Please refer to Appendix-\ref{appendix:proof_G_H_GF(p)}.
\end{IEEEproof}
\end{proposition}

Now, in order to make sure that both $\mathbf{G}$ and $\mathbf{H}$ are over GF($p$), it suffices to carefully design $\mathbf{U}$ and $\mathbf{U}'$, as illustrated in the next example.

\begin{example}
\label{example: q-ary_mds}
Let $p = 2$, and $\mathcal{J}$ represent an arbitrary subset of $\{0, 1, 2, \ldots, L-1\}$ that is closed under multiplication by $p$ modulo $L$. Define $\mathbf{U}=[\mathbf{u}_j]_{0 \leq j \leq L-1}$ to be the $J\times L$ full-rank matrix by deleting the last $L-J$ rows of $\mathbf{V}_L$, which is over GF($2^{m_L}$), and thus
\begin{equation}
\mathbf{G}=\mathbf{U}\widetilde{\mathbf{V}}_L=[\mathbf{I}_J~\mathbf{0}], %
\end{equation}
where $\mathbf{G}$ is over GF($2$). The local encoding kernels considered in \cite{x-tit} can be formulated as the $J\times J$ binary matrix in the form of $\mathbf{G}f(\mathbf{C}_L)\mathbf{G}^{\mathrm{T}}$. %
Now define $\mathbf{H} = \mathbf{V}_L[\mathbf{u}'_j]_{0\leq j \leq L-1}^{\mathrm{T}}$ with $[\mathbf{u}'_j]_{j\in \cal J}^{\mathrm{T}} = [\mathbf{u}_j]_{j\in {\cal J}}^{-1}$ and $[\mathbf{u}_j']_{j \notin \cal J} = \mathbf{0}$, so that $\mathbf{H}$ can be written as $[\mathbf{I}_J~\mathbf{A}]^\mathrm{T}$ for some $J\times(L-J)$ matrix $\mathbf{A}$. By Proposition \ref{prop: G H GF(p)}, $\mathbf{A}$ is over GF($2$). For instance, when $L = 9$ and $\mathcal{J} = \{1,2,4,5,7,8\}$, $\mathbf{H} = [\mathbf{I}_6~\mathbf{A}]^{\mathrm{T}}$ with $\mathbf{A} = [\mathbf{I}_3~\mathbf{I}_3]^{\mathrm{T}}$. \hfill $\blacksquare$
\end{example}

\begin{example}
\label{example: GH_construction}
As another design instance with $p = 2$, let $\mathbf{U}' = [\mathbf{u}'_j]_{0 \leq j \leq L-1}$ be the $J\times L$ full-rank matrix by deleting the first row as well as the last $L-J-1$ rows of $\mathbf{V}_L$, and define $\mathbf{H} = \mathbf{V}_L\mathbf{U}'^{\mathrm{T}}$. Because $\mathbf{V}_L^{2} = \begin{bmatrix}\begin{smallmatrix}
1&0 & \ldots & 0\\
0&0 & \ldots & 1\\
  \vdots & \vdots & \begin{sideways} $\ddots$ \end{sideways} &\vdots  \\
0&1&\ldots  &0
\end{smallmatrix}\end{bmatrix}$, one can show $\mathbf{H} = [\mathbf{0}~\mathbf{J}_{J}]^{\mathrm{T}}$, where $\mathbf{J}_J$ represents the $J\times J$ anti-diagonal identity matrix. %
Define $\mathbf{U} = [\mathbf{u}_j]_{0 \leq j \leq L-1}$ as
\begin{equation}
\label{eqn:U instance}
[\mathbf{u}_j]_{j\in \mathcal{J}} = ([\mathbf{u}'_j]^{-1}_{j\in \mathcal{J}})^{\mathrm{T}}, [\mathbf{u}_j]_{j\notin \mathcal{J}} = \mathbf{0}.
\end{equation}
For $\mathbf{G} = \mathbf{U}\widetilde{\mathbf{V}}_L$, it is over GF($2$) by Proposition \ref{prop: G H GF(p)}. %
As an illustration, let $L = 7$ and $\mathcal{J} = \{1, 2, 4\}$. One can readily obtain $\mathbf{G}=[\mathbf{A}~\mathbf{J}_3]$ with $\mathbf{A} =\begin{bmatrix} \begin{smallmatrix}1&1&0&1\\0&1&1&1\\1&1&1&0\end{smallmatrix}\end{bmatrix}$. %
\hfill $\blacksquare$
\end{example}

\begin{proposition}
\label{prop:GAH}
For arbitrary two polynomials $f_1(x)$ and $f_2(x)$ over GF($p$), the following equations hold
\begin{equation}
\label{eqn:GAH}
(\mathbf{G}f_1(\mathbf{C}_L)\mathbf{H})(\mathbf{G}f_2(\mathbf{C}_L)\mathbf{H}) = \mathbf{G}f_1(\mathbf{C}_L)f_2(\mathbf{C}_L)\mathbf{H},
\end{equation}
\begin{equation}
\label{eqn:HAG}
(\mathbf{H}^{\mathrm{T}}f_1(\mathbf{C}_L)\mathbf{G}^{\mathrm{T}})(\mathbf{H}^{\mathrm{T}}f_2(\mathbf{C}_L)\mathbf{G}^{\mathrm{T}}) = \mathbf{H}^{\mathrm{T}}f_1(\mathbf{C}_L)f_2(\mathbf{C}_L)\mathbf{G}^{\mathrm{T}}.
\end{equation}
\begin{IEEEproof}
Please refer to Appendix-\ref{appendix:proof_GAH}.
\end{IEEEproof}
\end{proposition}

Now assume $\mathbf{G}$ and $\mathbf{H}$ are constructed subject to \eqref{U'_def}-\eqref{eqn:GH_def}, and they are both matrices over GF($p$), which can be satisfied based on the condition in Proposition \ref{prop: G H GF(p)}. The next corollary is a direct consequence of Proposition \ref{prop:GAH}. 

\begin{corollary}
\label{cor: multiplication_matrices_closed}
In the definition of $\mathcal{C}$ in \eqref{eqn:circular-shift-based_kernel}, when $\mathbf{P}$ and $\mathbf{Q}$ are respectively set to $\mathbf{G}$ and $\mathbf{H}$, or $\mathbf{H}^\mathrm{T}$ and $\mathbf{G}^\mathrm{T}$, then 
$\mathcal{C}$ is closed under multiplication by matrices in it. 
\end{corollary}

\begin{figure}[!h]
\centering
\scalebox{0.7}
{\includegraphics{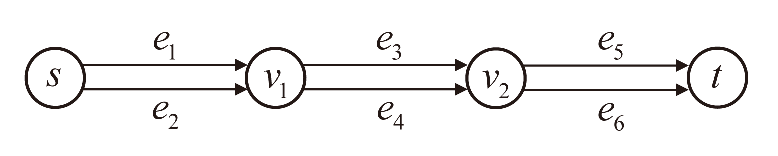}}
\caption{A network consists of a unique source node, $2$ intermediate nodes and a single receiver. It is used in Example \ref{example: GAH_recoding} and Example \ref{example: GAH_decoding}. }
\label{Fig:4_nodes_network}
\end{figure}

\begin{example}
\label{example: GAH_recoding}
Consider the four node network shown in Fig. \ref{Fig:4_nodes_network}, which consists of a source node $s$, two intermediate nodes $v_1$ and $v_2$, and a receiver $t$. Let $p = 2$, $L = 5$ and $\mathcal{J} = \{1, 2, 3, 4\}$, so that $m_L = 4$ and $J = |\mathcal{J}| = 4$. %
Respectively define the $4\times 5$ binary matrix $\mathbf{G}=[\mathbf{I}_4~\mathbf{0}]$ and the $5\times 4$ binary matrix $\mathbf{H}=[\mathbf{I}_4~\mathbf{1}]^\mathrm{T}$. %
The source node $s$ thus generates two 4-dimensional binary data units $\mathbf{m}_1 = [m_{11} ~ m_{12} ~ m_{13} ~ m_{14}]$ and $\mathbf{m}_2 = [m_{21} ~ m_{22} ~ m_{23} ~ m_{24}]$. %
The data units transmitted on the two outgoing edges $e_1$, $e_2$ of s are, respectively,  $\mathbf{m}_{e_1} = \mathbf{m}_1 = [m_{11} ~ m_{12} ~ m_{13} ~ m_{14}]$ and $\mathbf{m}_{e_2} = \mathbf{m}_2 = [m_{21} ~ m_{22} ~ m_{23} ~ m_{24}]$. %
The local encoding kernels at $v_1$ and $v_2$, which are all $4 \times 4$ binary matrices, are 
\begin{equation}
\begin{split}
&\mathbf{K}_{e_1, e_4} = \mathbf{K}_{e_2, e_3} = \mathbf{K}_{e_3, e_6} = \mathbf{K}_{e_4, e_5} = \mathbf{0}, \\
&\mathbf{K}_{e_1, e_3} = \mathbf{G}\mathbf{C}_5\mathbf{H}, ~ \mathbf{K}_{e_2, e_4} = \mathbf{G}\mathbf{C}_{5}^{2}\mathbf{H}, \\
&\mathbf{K}_{e_3, e_5} = \mathbf{G}\mathbf{C}_{5}^{3}\mathbf{H}, ~ \mathbf{K}_{e_4, e_6} = \mathbf{G}\mathbf{C}_{5}^{4}\mathbf{H}.
\end{split}
\end{equation}
According to \eqref{eqn:GAH}, the binary data units transmitted on edges $e_5$ and $e_6$, can be computed as
\begin{equation*}\begin{split}
\mathbf{m}_{e_5} &= \mathbf{m}_{e_1} (\mathbf{G}\mathbf{C}_5\mathbf{H}) (\mathbf{G}\mathbf{C}_{5}^{3}\mathbf{H}) = \mathbf{m}_{e_1} \mathbf{G}\mathbf{C}_{5}^{4}\mathbf{H} \\
&= [m_{11}+m_{12}~m_{11}+m_{13}~m_{11}+m_{14}~m_{11}], \\
\mathbf{m}_{e_6} &= \mathbf{m}_{e_2}(\mathbf{G}\mathbf{C}_{5}^{2}\mathbf{H}) (\mathbf{G}\mathbf{C}_{5}^{4}\mathbf{H}) = \mathbf{m}_{e_2}\mathbf{G}\mathbf{C}_5\mathbf{H} \\
&= [m_{24}~m_{21}+m_{24}~m_{22}+m_{24}~m_{23}+m_{24}]. 
\end{split}
\end{equation*}
\hfill $\blacksquare$
\end{example}

\begin{corollary}
\label{cor: GAH_I}
Consider a polynomial $f(x)$ over GF($p$) subject to $f(\beta^j) \neq 0$ for all $j \in \mathcal{J}$. Let $d(x) = f(x)^{p^{m_L}-2}$, so that the following equations hold
\begin{equation}
\label{eqn: GAHGDH = I}
(\mathbf{G}f(\mathbf{C}_L)\mathbf{H})(\mathbf{G}d(\mathbf{C}_L)\mathbf{H}) = \mathbf{I}_J,
\end{equation}
\begin{equation}
\label{eqn: H'AG'H'DG' = I}
(\mathbf{H}^\mathrm{T}f(\mathbf{C}_L)\mathbf{G}^\mathrm{T})(\mathbf{H}^\mathrm{T}d(\mathbf{C}_L)\mathbf{G}^\mathrm{T}) = \mathbf{I}_J.
\end{equation}
\begin{IEEEproof}
Please refer to Appendix-\ref{appendix:proof_GAH_I}.
\end{IEEEproof}
\end{corollary}

The above corollary asserts that in the definition of $\mathcal{C}$, when  $\mathbf{P} = \mathbf{G}$, $\mathbf{Q} = \mathbf{H}$ or $\mathbf{P} = \mathbf{H}^\mathrm{T}$, $\mathbf{Q} = \mathbf{G}^\mathrm{T}$, every nonzero matrix in $\mathcal{C}$ is invertible. 

It is worthwhile noticing that in the formulation of circular-shift LNC \cite{Tang_LNC_TIT}-\cite{Tang_CL19_encoding_decoding_circular_shift_LNC}, matrices $\mathbf{G}$ and $\mathbf{H}$ similar to the ones defined in \eqref{U'_def}-\eqref{eqn:GH_def} have been exploited to characterize the encoding and the decoding process. %
However, $\mathbf{G}$ and $\mathbf{H}$ were not adopted to design local encoding kernels, so that the circular-shift linear network codes considered in \cite{Tang_LNC_TIT}-\cite{Tang_CL19_encoding_decoding_circular_shift_LNC} have rate less than $1$, that is, the data units transmitted along the network contain redundant symbols. %
A key reason for this is the absence of the nontrivial properties established in \eqref{eqn:GAH}, \eqref{eqn:HAG}, \eqref{eqn: GAHGDH = I} and \eqref{eqn: H'AG'H'DG' = I}. In addition, $\mathbf{G}$ and $\mathbf{H}$ investigated in this section are more general than the ones considered in \cite{Tang_LNC_TIT}-\cite{Tang_CL19_encoding_decoding_circular_shift_LNC}, 
because in the construction of $\mathbf{G}$ and $\mathbf{H}^\mathrm{T}$ in \cite{Tang_LNC_TIT}-\cite{Tang_CL19_encoding_decoding_circular_shift_LNC}, $\mathcal{J}$ is typically selected to be the integers between $1$ and $L-1$ that are co-prime with $L$, while $\mathcal{J}$ is only required to be closed under multiplication by $p$ modulo $L$ in the present consideration. %
For instance, when $L = 15$ and $p = 2$, $\mathcal{J}$ is fixed to $\{1,2,4,7,8,11,13,14\}$ with $J = |\mathcal{J}| = 8$, while $\mathcal{J}$ can also be set to $\{1,2,3,4,6,7,8,9,11,12,13,14\}$ with $J = |\mathcal{J}| = 12$ in the more general construction.



\subsection{Construction of circular-shift-based vector linear solutions}
\label{subsec: GH_linear_solution}

In this subsection, assume $\mathbf{G}$ and $\mathbf{H}$ are constructed subject to \eqref{U'_def}-\eqref{eqn:GH_def}, and they are both matrices over GF($p$). %
Let $\mathbf{P}$ and $\mathbf{Q}$ in \eqref{eqn:circular-shift-based_kernel} respectively equal to $\mathbf{G}$ and $\mathbf{H}$, or $\mathbf{H}^\mathrm{T}$ and $\mathbf{G}^\mathrm{T}$. %
In this way, $\mathcal{C}$ is closed under multiplication by matrices in it, and every nonzero matrix in $\mathcal{C}$ is invertible. %

We shall establish a connection between an $L$-dimensional circular-shift linear solution at rate $J/L$ and a $J$-dimensional circular-shift-based vector linear solution at rate $1$. We first review the formulation of circular-shift LNC at rate $J/L < 1$. 


For $L$-dimensional circular-shift LNC at rate $J/L$, the source node $s$ generates $\omega$ source data units $\mathbf{m}'_1, \mathbf{m}'_2, \ldots, \mathbf{m}'_\omega$, each of which is a $J$-dimensional row vector over GF($p$), while the date unit $\mathbf{m}_e$ transmitted along every edge $e$ is an $L$-dimensional row vector over GF($p$). In particular, the data units $\mathbf{m}_e$, $e \in \mathrm{Out}(s)$, transmitted on the outgoing edges of $s$ are computed according to \eqref{eqn: circular-shift_LNC_encoding_matrix} based on $\mathbf{m}'_1, \mathbf{m}'_2, \ldots, \mathbf{m}'_\omega$ and a $\omega J \times \omega L$ source encoding matrix $\mathbf{G}_s$ over GF($p$). %
Throughout this paper, as long as $\mathbf{P}$ is defined, $\mathbf{G}_s$ used in an $L$-dimensional circular-shift linear network code at rate $J/L$ is set to $\mathbf{P}_{\omega \otimes}$, that is, \begin{equation}
\mathbf{G}_s = \mathbf{P}_{\omega \otimes} =\mathbf{I}_{\omega} \otimes \mathbf{P}.
\end{equation}
In turns out that in $L$-dimensional circular-shift LNC at rate $J/L$, every data unit transmitted along the network carries $L-J$ redundant $p$-ary data symbols.

Recall that the local encoding kernels $\mathbf{K}_{d,e}$ in an $L$-dimensional circular-shift linear network code ($\mathbf{K}_{d, e}$) over GF($p$) are selected from the set $\left\{\sum\nolimits_{j=0}^{L-1}a_j\mathbf{C}_L^j: a_j \in \mbox{GF($p$)} \right\}$. %
Thus, $(\mathbf{K}_{d, e})$ naturally induces a $J$-dimensional circular-shift-based vector linear network code $(\mathbf{P}\mathbf{K}_{d, e}\mathbf{Q})$, in which the $J\times J$ local encoding kernel for adjacent pair $(d,e)$ is $\mathbf{P}\mathbf{K}_{d, e}\mathbf{Q}$.

For an $L$-dimensional circular-shift linear network code $(\mathbf{K}_{d,e})$ at rate $J/L$, the global encoding kernel $\mathbf{F}_e$ for every edge $e$ is still an $\omega L \times L$ matrix over GF($p$). The data unit $\mathbf{m}_e$ transmitted on edge $e$ can be expressed as
\begin{equation}
\mathbf{m}_e = [\mathbf{m}'_i]_{1 \leq i \leq \omega}
\mathbf{P}_{\omega \otimes}\mathbf{F}_e.
\end{equation}
The code $(\mathbf{K}_{d,e})$ is qualified to be a linear solution at rate $J/L$ if for every receiver $t$, the $\omega J\times \omega L$ matrix $\mathbf{P}_{\omega \otimes}[\mathbf{F}_e]_{e \in \mathrm{In(t)}}$ over GF($p$) has full rank $\omega J$. %
In contrast, for the induced $J$-dimensional circular-shift-based vector linear network code $(\mathbf{P}\mathbf{K}_{d, e}\mathbf{Q})$, based on Proposition \ref{prop:GAH}, one can show that the data unit transmitted on every edge $e$, to be denoted by 
$\bar{\mathbf{m}}_e$, can be characterized as $\bar{\mathbf{m}}_e = [\mathbf{m}'_i]_{1 \leq i \leq \omega}\mathbf{P}_{\omega \otimes}\mathbf{F}_e\mathbf{Q}$. %
Different from $\mathbf{m}_e$, $\bar{\mathbf{m}}_e$ is a $J$-dimensional row vector that does not contain any redundant $p$-ary data symbols. %
The code $(\mathbf{P}\mathbf{K}_{d, e}\mathbf{Q})$ is qualified to be a linear solution (at rate $1$) if for every receiver $t$, the $\omega J\times \omega J$ matrix $\mathbf{P}_{\omega \otimes}[\mathbf{F}_e]_{e \in \mathrm{In(t)}}\mathbf{Q}_{\omega \otimes}$ over GF($p$) has full rank $\omega J$.

\begin{theorem}
\label{theorem: circular_shift_solutions_to_circular_shift_based_solution}
Consider an $L$-dimensional circular-shift linear network code $(\mathbf{K}_{d,e})$ at rate $J/L$, and the $J$-dimensional circular-shift-based vector linear network code $(\mathbf{P}\mathbf{K}_{d,e}\mathbf{Q})$ induced from $(\mathbf{K}_{d,e})$. The code $(\mathbf{K}_{d,e})$ equipped with the source encoding matrix $\mathbf{P}_{\omega \otimes}$ is a linear solution at rate $J/L$ if and only if $(\mathbf{P}\mathbf{K}_{d,e}\mathbf{Q})$ is a linear solution (at rate $1$). 
\begin{IEEEproof}
Please refer to Appendix-\ref{appendix: theorem: circular_shift_solutions_to_circular_shift_based_solution}.
\end{IEEEproof}
\end{theorem}

Compared with circular-shift LNC, the most significant advantage of circular-shift-based vector LNC is that the data units transmitted in a network contain no redundancy. %
We next analyze the computational complexity of generating a data unit at an intermediate node for circular-shift LNC and for circular-shift-based vector LNC. %
Let $\mathbf{K}_1 = \sum_{j = 1}^{\delta_1} a_{1,j}\mathbf{C}_L^{l_{1,j}}$, $\mathbf{K}_2 = \sum_{j = 1}^{\delta_2} a_{2,j}\mathbf{C}_L^{l_{2,j}}$ be two circulant matrices over $\mathrm{GF}(p)$. %
For an $L$-dimensional row vector $\mathbf{m}_i$, $i \in {1, 2}$ over GF($p$), computing $\mathbf{m}_i\mathbf{K}_i$ takes $(\delta_i-1)L$ additions and $\delta_i L$ multiplications over GF($p$). %
Consequently, it takes $(\delta_1+\delta_2-1)L$ additions and $(\delta_1+\delta_2)L$ multiplications over GF($p$) to compute $\mathbf{m}_1\mathbf{K}_1+\mathbf{m}_2\mathbf{K}_2$. %
Let $\mathbf{G} = [\mathbf{I}_J ~ \mathbf{0}]$ and $\mathbf{H} = [\mathbf{I}_J ~ \mathbf{A}]^\mathrm{T}$ over GF($p$) subject to conditions \eqref{U'_def}-\eqref{eqn:GH_def}, and let $h$ denote the number of nonzero elements in $\mathbf{A}$. %
For a $J$-dimensional row vector $\bar{\mathbf{m}}_i$, $i \in {1, 2}$ over GF($p$), computing $\bar{\mathbf{m}}_i(\mathbf{G}\mathbf{K}_i\mathbf{H})$, which is equivalent to computing $\left(\sum_{j = 1}^\delta \bar{\mathbf{m}}_i\mathbf{G}(a_j\mathbf{C}_L^{j})\right)\mathbf{H}$, takes $(\delta_i-1)J + h$ additions and $\delta_i J + h$ multiplications over GF($p$). %
Consequently, it takes $(\delta_1+\delta_2-1)J + h$ additions and $(\delta_1+\delta_2)J + h$ multiplications over GF($p$) to compute $\bar{\mathbf{m}}_1(\mathbf{G}\mathbf{K}_1\mathbf{H}) + \bar{\mathbf{m}}_2(\mathbf{G}\mathbf{K}_1\mathbf{H}) = (\bar{\mathbf{m}}_1\mathbf{G}\mathbf{K}_1+\bar{\mathbf{m}}_2\mathbf{G}\mathbf{K}_2)\mathbf{H}$. %
When $p = 2$, no multiplication is required for both cases. %
It turns out that the coding complexity of circular-shift-based vector LNC is comparable to and, in some cases, even lower than that of circular-shift LNC. For instance, let $p = 2$, $L = 9$, and $J = \phi(L) = 6$. As illustrated in Example \ref{example: q-ary_mds}, $\mathbf{H}$ can be set as $\mathbf{H} = [\mathbf{I}_J ~ \mathbf{A}]^\mathrm{T}$ with $\mathbf{A} = [\mathbf{I}_3~\mathbf{I}_3]^{\mathrm{T}}$, so that $h = 6$. Thus, when $\delta_1 + \delta_2 > 3$, the number of XORs required to compute $\bar{\mathbf{m}}_1(\mathbf{G}\mathbf{K}_1\mathbf{H}) + \bar{\mathbf{m}}_2(\mathbf{G}\mathbf{K}_1\mathbf{H})$ in circular-shift-based vector LNC is smaller than that required to compute $\mathbf{m}_1\mathbf{K}_1+\mathbf{m}_2\mathbf{K}_2$ in circular-shift LNC. 

Next, consider the special case that $\mathcal{J} = \{1 \leq j \leq L-1: \gcd(j, L) = 1\}$ and $J = \phi(L)$. %
Recall that Algorithm 8 in \cite{Tang_Sun_Circular-shift_LNC_TCOM} showed that when $\frac{m_L}{\phi(L)}2^{m_L} > |T|$, a circular-shift linear solution over GF($2$) at rate $\phi(L)/L$ can be efficiently constructed. %
The algorithm can be extended to efficiently construct a circular-shift linear solution over GF($p$) at rate $\phi(L)/L$ in a straightforward way. %
According to Theorem \ref{theorem: circular_shift_solutions_to_circular_shift_based_solution}, we assert that a $J$-dimensional circular-shift-based vector linear solution over GF($p$) can also be efficiently constructed.
\begin{corollary}
\label{cor: efficient_construction_circular_shift_based}
When $\frac{m_L}{\phi(L)}p^{m_L} > |T|$, a $\phi(L)$-dimensional circular-shift-based vector linear solution over GF($p$) (at rate $1$) can be efficiently constructed.
\end{corollary}

We now explicitly characterize the decoding matrix at a receiver $t$, which is the inverse matrix of $\mathbf{P}_{\omega \otimes}[\mathbf{F}_e]_{e \in \mathrm{In}(t)}\mathbf{Q}_{\omega \otimes}$. %
Let $\mathbf{\Psi}(x)$ be such an $\omega \times \omega$ matrix in which every entry is a polynomial over GF($p$) that $\mathbf{\Psi}(\mathbf{C}_L) = [\mathbf{F}_e]_{e \in \mathrm{In}(t)}$. %
Meanwhile, $\mathbf{\Psi}(\beta^j)$ is a matrix over GF($p^{m_L}$), where $\beta$ is a primitive $L^{\mathrm{th}}$ root of unity over GF($p$). According to the following rank equation 
\begin{equation}
\label{eqn:rank_equation}
\mathrm{rank}(\mathbf{\Psi}(\mathbf{C}_L)) = \sum\nolimits_{j = 0}^{L-1} \mathrm{rank}(\mathbf{\Psi}(\beta^j)),
\end{equation}
which was proved in \cite{Tang_Sun_Circular-shift_LNC_TCOM} (Theorem 1 therein) to construct circular-shift linear solutions (at rate smaller than $1$) over an arbitrary multicast network. 
\footnote{In \cite{Tang_Sun_Circular-shift_LNC_TCOM}, only the case $p = 2$ and $L$ an odd integer is considered. However, an essentially same approach can be adopted to prove \eqref{eqn:rank_equation} for more general prime $p$ and an integer $L$ co-prime with $p$. The proof details are omitted here.}  %
The next lemma further asserts the full rank of $\mathbf{G}_{\omega \otimes}\mathbf{\Psi}(\mathbf{C}_L)\mathbf{H}_{\omega \otimes}$ and $\mathbf{H}_{\omega \otimes}^\mathrm{T}\mathbf{\Psi}(\mathbf{C}_L)\mathbf{G}_{\omega \otimes}^\mathrm{T}$, which has been utilized in proving the sufficiency part in Theorem \ref{theorem: circular_shift_solutions_to_circular_shift_based_solution}. 

\begin{lemma}
\label{lemma: decoding_matrix_for_GAH}
Assume $\mathrm{rank}(\mathbf{\Psi}(\beta^j)) = \omega$ for all $j \in \mathcal{J}$. Define the counterpart $\omega \times \omega$ matrix %
\begin{equation}
\label{eqn: Psi_inverse_def}
\mathbf{\Psi}'(x) = \mathrm{det}(\mathbf{\Psi}(x))^{p^{m_L}-2} \mathrm{Adj}(\mathbf{\Psi}(x)),
\end{equation}
where $\mathrm{det}(\mathbf{\Psi}(x))$ represents the polynomial over GF($p$) obtained by calculating the determinant of $\mathbf{\Psi}(x)$, and $\mathrm{Adj}(\cdot)$ denotes the adjugate of a matrix. %
We have, 
\begin{equation}
\label{eqn: decoding_matrix_for_GAH}
(\mathbf{G}_{\omega \otimes}\mathbf{\Psi}(\mathbf{C}_L)\mathbf{H}_{\omega \otimes}) (\mathbf{G}_{\omega \otimes}\mathbf{\Psi}'(\mathbf{C}_L)\mathbf{H}_{\omega \otimes}) = \mathbf{I}_{\omega J},
\end{equation}
\begin{equation}
\label{eqn: decoding_matrix_for_HTAGT}
(\mathbf{H}_{\omega \otimes}^\mathrm{T}\mathbf{\Psi}(\mathbf{C}_L)\mathbf{G}_{\omega \otimes}^\mathrm{T}) (\mathbf{H}_{\omega \otimes}^\mathrm{T}\mathbf{\Psi}'(\mathbf{C}_L)\mathbf{G}_{\omega \otimes}^\mathrm{T}) = \mathbf{I}_{\omega J},
\end{equation}
\begin{IEEEproof}
Please refer to Appendix-\ref{appendix: proof_GAH_decoding_matrix}.
\end{IEEEproof}
\end{lemma}

\begin{example}
\label{example: GAH_decoding}
Recall the network in Example \ref{example: GAH_recoding}. %
For the receiver $t$, the $8\times 8$ binary matrix $[\mathbf{F}_e]_{e \in \mathrm{In(t)}}$ is 
\begin{equation*}
[\mathbf{F}_e]_{e \in \mathrm{In}(t)} = [\mathbf{F}_{e_5} ~ \mathbf{F}_{e_6}] 
= \begin{bmatrix}
\mathbf{G}\mathbf{C}_{5}^{4}\mathbf{H} & \mathbf{0} \\
\mathbf{0} & \mathbf{G}\mathbf{C}_5\mathbf{H}
\end{bmatrix}.
\end{equation*}
According to Lemma \ref{lemma: decoding_matrix_for_GAH}, we can obtain the decoding matrix 
\begin{equation*}
\label{eqn: example decoding matrix}
\mathbf{D}_t = 
 \begin{bmatrix}
\mathbf{G}\mathbf{C}_5\mathbf{H} & \mathbf{0} \\
\mathbf{0} & \mathbf{G}\mathbf{C}_{5}^{4}\mathbf{H}
\end{bmatrix}.
\end{equation*}
Subsequently, based on $\mathbf{D}_t$, the source data units can be recovered, that is, 
\[
[\mathbf{m}_{e_5} ~ \mathbf{m}_{e_6}]\mathbf{D}_t = [m_{11} ~ m_{12} ~ m_{13} ~ m_{14} ~ m_{21} ~ m_{22} ~ m_{23} ~ m_{24}].
\]
\hfill $\blacksquare$
\end{example}

\subsection{Circular-shift-based vector LNC on the Combination Network}
\label{subsec: combination_network}
\begin{figure}[!h]
\centering
\scalebox{0.55}
{\includegraphics{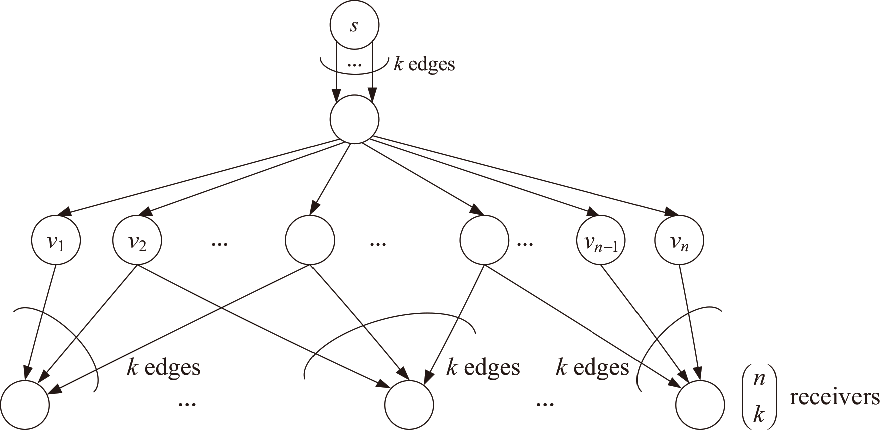}}
\caption{The classical $(n,k)$-Combination Network with four layers.}
\label{Fig:Combination_Network}
\end{figure}

Consider the classical $(n, k)$-Combination Network depicted in Fig. \ref{Fig:Combination_Network}, where the unique source node $s$ generates $k$ data units $\mathbf{m}_1, \ldots, \mathbf{m}_k$ to be transmitted to each of the $\binom{n}{k}$ receivers, and every edge has unit capacity to transmit a data unit. %
For a $L$-dimensional vector linear network code, it prescribes a $kL \times L$ matrix $\mathbf{F}_{e_j} = [\mathbf{K}_{e_{j,1}}^\mathrm{T}~\ldots~\mathbf{K}_{e_{j,k}}^\mathrm{T}]^\mathrm{T}$ consisting of $k$ local encoding kernels for the incoming edge $e_j$ of every layer-$3$ node $v_j$, %
so that the data unit $\mathbf{m}_{e_j}$ transmitted on $e_j$ can be interpreted as %
$\mathbf{m}_{e_j} = [\mathbf{m}_1~\ldots~\mathbf{m}_k]\mathbf{F}_{e_j} = \sum\nolimits_{1\leq i \leq k}\mathbf{m}_i\mathbf{K}_{e_{j,i}}$. %
An $L$-dimensional vector linear network code  is qualified as a linear solution if for every receiver connected with $k$ layer-$3$ nodes $v_{l_1}, \ldots v_{l_k}$, the $kL\times kL$ matrix $[\mathbf{F}_{e_{l_j}}]_{1\leq j \leq k}$ has full rank $kL$. %
In the literature, the Combination Network is one of the most celebrated network topologies and it has been widely studied in various topics (See, e.g., \cite{sun2016vector}-\cite{han2020tit},\cite{tang2022multicast}). %
In this subsection, we particularly investigate circular-shift-based vector LNC on the $(n, k)$-Combination Network, because circular-shift-based vector linear solutions on the $(n, k)$-Combination Network can be applied to characterize and design MDS array codes, as to be clear in the next section.

First, according to Theorem \ref{theorem: circular_shift_solutions_to_circular_shift_based_solution}, if we adopt the same setting of $\mathbf{P} = \mathbf{G}$ and $\mathbf{Q} = \mathbf{H}$ as in the previous subsection, %
then an arbitrary $L$-dimensional circular-shift linear solution $(\mathbf{K}_{d,e})$ at rate $J/L$ on the $(n, k)$-Combination Network induces a $J$-dimensional circular-shift-based vector linear solution $(\mathbf{P}\mathbf{K}_{d,e}\mathbf{Q})$. %

One may notice that the $(n, k)$-Combination Network only contains one intermediate node with multiple incoming edges, so this is the only node required to perform network coding. Consequently, recoding processes are no longer needed on the Combination Network, which allows us to have more flexible design of matrices $\mathbf{P}$ and $\mathbf{Q}$ over GF($p$) in \eqref{eqn:circular-shift-based_kernel} to formulate local encoding kernels of circular-shift-based vector LNC. %
In the remaining part of this subsection, we first introduce an alternative way to design matrices $\mathbf{P}$ and $\mathbf{Q}$ over GF($p$), so that a new $J$-dimensional circular-shift-based vector linear solution $(\mathbf{P}\mathbf{K}_{d,e}\tau(\mathbf{C}_L)\mathbf{Q})$ can be induced based on an arbitrary $L$-dimensional circular-shift linear solution $(\mathbf{K}_{d,e})$ at rate $J/L$ on the $(n, k)$-Combination Network. Herein, $\tau(x)$ denotes the following polynomial over GF($p$) of degree $L - J$
\begin{equation}
\label{def: tau(x)}
\tau(x)=\prod\nolimits_{j \notin \cal J}(x-\beta^j).
\end{equation}
Recall that as defined in Section \ref{subsec:Mathematical notations and useful lemmas}, $\mathcal{J}$ is a subset of $\{0, 1, 2, \ldots, L-1\}$ that is closed under multiplication by $p$ modulo $L$, and $J = |\mathcal{J}|$. %

Same as the definition of $\mathbf{U}$ and $\mathbf{G}$ in Section \ref{subsec:GH}, assume 
\begin{equation}
\label{U_bar_J}
\mathrm{rank}([\mathbf{u}_j]_{j\in \mathcal{J}}) = J,~\mathrm{rank}([\mathbf{u}_j]_{j\in \bar{\mathcal{J}}}) = J,~\mathbf{G} = \mathbf{U}\widetilde{\mathbf{V}}_L,
\end{equation}
where ${\bar{\mathcal{J}}}=\{L-j: j\in \cal J \}$ so that $|\bar{\mathcal{J}}| = J$. %
Analogous to the construction \eqref{U'_def}-\eqref{eqn:GH_def} in the previous subsection, 
we define the $L\times J$ matrices $\mathbf{H}$ and $\bar{\mathbf{H}}$ over GF($p$) as follows
\begin{equation}
\label{eqn:H_H_bar_def}
\mathbf{H}=\mathbf{V}_L\mathbf{U}'^{\mathrm{T}}, \bar{\mathbf{H}} = \mathbf{V}_L\mathbf{U}''^{\mathrm{T}},
\end{equation}
where $\mathbf{U}' = [\mathbf{u}'_j]_{0\leq j\leq L-1}$ and $\mathbf{U}'' = [\mathbf{u}''_j]_{0\leq j\leq L-1}$ are $J\times L$  matrices over GF($p^{m_L}$) obtained by 
\begin{align}
\label{U' new_def}
[\mathbf{u}'_j]_{j\in \cal J}^{\mathrm{T}} = L_{\mathrm{GF}(p)}^{-1}[\mathbf{u}_j]_{j\in {\cal J}}^{-1},~~ \mathbf{u}'_j=\mathbf{0}~\forall j \notin  {\cal J}, \\
\label{U'' new _def}
[\mathbf{u}''_j]_{j\in \bar{\mathcal{ J}}}^{\mathrm{T}} = L_{\mathrm{GF}(p)}^{-1}[\mathbf{u}_j]_{j\in \bar{\mathcal{ J}}}^{-1},~~ \mathbf{u}''_j=\mathbf{0}~\forall j \notin  \bar{\cal J}.
\end{align}
By \eqref{U' new_def}, $\mathbf{G}\mathbf{H} = \mathbf{I}_J$. %
Notice that $\mathbf{H}=\bar{\mathbf{H}}$ for the special case ${\cal J}={\bar{\mathcal{J}}}$. Similar to the case considered in the previous subsection, the matrices $\mathbf{G}$, $\mathbf{H}$ and $\bar{\mathbf{H}}$ are not necessarily over GF($p$), but we can show the following.

\begin{proposition}
\label{prop: H H-bar over GF(p)}
If $\mathbf{G}$ is over GF($p$), then the counterpart $L\times J$ matrices $\mathbf{H}$ and $\bar{\mathbf{H}}$ defined in \eqref{eqn:H_H_bar_def} are also over GF($p$).
\begin{IEEEproof}
Since $\mathbf{G}$ is over GF($p$) and $[\mathbf{u}']_{j\notin \mathcal{J}} =\mathbf{0}$, Proposition \ref{prop: G H GF(p)} directly asserts that $\mathbf{H}$ is over GF($p$). %
Since ${\bar{\mathcal{J}}}$ is also closed under multiplication by $p$ modulo $L$, the assumption that $\mathbf{G}$ is over GF($p$) and $[\mathbf{u}'']_{j\notin \bar{\mathcal{J}}} =\mathbf{0}$ guarantees that $\bar{\mathbf{H}} = \mathbf{V}_L\mathbf{U}''^{\mathrm{T}}$ is over GF($p$) according to Proposition \ref{prop: G H GF(p)}.
\end{IEEEproof}
\end{proposition}

\begin{example}
Let $p = 2$, and $\mathcal{J}$ represent an arbitrary subset of $\{0, 1, 2, \ldots, L-1\}$ that is closed under multiplication by $p$ modulo $L$. In addition, $\bar{\mathcal{J}} = \{L-j: j \in \mathcal{J}\}$.
Define $\mathbf{U}=[\mathbf{u}_j]_{0 \leq j \leq L-1}$ to be the $J\times L$ full-rank matrix by deleting the last $L-J$ rows of $\mathbf{V}_L$, and thus
\begin{equation}
\mathbf{G}=\mathbf{U}\widetilde{\mathbf{V}}_L=[\mathbf{I}_J~\mathbf{0}].
\end{equation}
The $L\times J$ matrices $\mathbf{H}$ and $\bar{\mathbf{H}}$ constructed via \eqref{eqn:H_H_bar_def}-\eqref{U'' new _def} can be respectively written as $[\mathbf{I}_J~\mathbf{A}]^\mathrm{T}$ and $[\mathbf{I}_J~\bar{\mathbf{A}}]^\mathrm{T}$. %
By Proposition \ref{prop: H H-bar over GF(p)}, both $\mathbf{A}$ and $\bar{\mathbf{A}}$ are over GF($p$). %
For instance. When $L = 7$, $\mathcal{J} = \{1,2,4\}$ and ${\bar{\mathcal{J}}} = \{3,5,6\}$, %
we have $\mathbf{A} = \begin{bmatrix}\begin{smallmatrix}
1&0&1&1\\1 &1& 1& 0\\0 &1& 1& 1
\end{smallmatrix}\end{bmatrix}$ and
$\bar{\mathbf{A}} = \begin{bmatrix}\begin{smallmatrix}
1&1&1&0\\0 &1& 1& 1\\1 &1& 0& 1
\end{smallmatrix}\end{bmatrix}$.
\hfill $\blacksquare$
\end{example}

Analogous to Proposition \ref{prop:GAH} in the previous subsection, which is useful in the proof of 
Theorem \ref{theorem: circular_shift_solutions_to_circular_shift_based_solution} and Lemma \ref{lemma: decoding_matrix_for_GAH}, 
the next proposition will be useful in the proof of Theorem \ref{theorem: circular_shift_solutions_to_circular_shift_based_solution_GG} 
and Lemma \ref{lemma: decoding_matrix_for_GAGT} in the sequel. 
\begin{proposition}
\label{prop: GkG'H''k'H=I}
For arbitrary two polynomials $f_1(x)$ and $f_2(x)$ over GF($p$), the following equations hold
\begin{equation}
\label{eqn:GAG'}
\begin{split}
&(\mathbf{G}f_1(\mathbf{C}_L)\tau(\mathbf{C}_L)\mathbf{G}^\mathrm{T})(\bar{\mathbf{H}}{^\mathrm{T}}f_2(\mathbf{C}_L)\mathbf{H})\\
=&\mathbf{G}f_1(\mathbf{C}_L)\tau(\mathbf{C}_L)f_2(\mathbf{C}_L)\mathbf{H}.
\end{split}
\end{equation}
\begin{IEEEproof}
Please refer to Appendix-\ref{appendix:proof_GkG'H''k'H_I}.
\end{IEEEproof}
\end{proposition}

Assume $\mathbf{G}$ and $\mathbf{H}$ are constructed subject to \eqref{U_bar_J} and \eqref{eqn:H_H_bar_def}, and they are both matrices over GF($p$). %
Notice that $\mathbf{G}$ and $\mathbf{H}$ also satisfy \eqref{U'_def}-\eqref{eqn:GH_def}. %
Let $\mathbf{P}$ and $\mathbf{Q}$ in \eqref{eqn:circular-shift-based_kernel} respectively equal to $\mathbf{G}$ and $\mathbf{H}$. %
Notice that an arbitrary $L$-dimensional circular-shift linear network code $(\mathbf{K}_{d,e})$ not only induces a $J$-dimensional circular-shift-based vector linear network code ($\mathbf{P}\mathbf{K}_{d, e}\mathbf{Q}$), but also induces another $J$-dimensional circular-shift-based vector linear network code ($\mathbf{P}\mathbf{K}_{d, e}\mathbf{Q}'$), where 
\begin{equation}
\mathbf{Q}' = \tau(\mathbf{C}_L)\mathbf{G}^\mathrm{T}
\end{equation}
with $\tau(x)$ defined in \eqref{def: tau(x)}. %

Consider an $L$-dimensional circular-shift linear network code $(\mathbf{K}_{d,e})$. For edge $e_j$ incoming to every layer-$3$ node $v_j$, $1 \leq j \leq n$, let $\mathbf{F}_{e_{j}}$ denote its global encoding kernel, which is a $kL\times L$ matrix over GF($p$). %
For the $J$-dimensional circular-shift-based vector linear network code $(\mathbf{P}\mathbf{K}_{d, e}\mathbf{Q}')$ induced from ($\mathbf{K}_{d, e}$), the $kJ \times J$ global encoding kernel for edge $e_j$ can be expressed as $\mathbf{P}_{k \otimes}\mathbf{F}_{e_{j}}\mathbf{Q}'$. %
Since every receiver is connected with $k$ layer-$3$ nodes $v_{l_1}, \ldots v_{l_k}$, $(\mathbf{P}\mathbf{K}_{d, e}\mathbf{Q}')$ qualifies as a linear solution if for every receiver, 
the $kJ\times kJ$ matrix $\mathbf{P}_{k \otimes}[\mathbf{F}_{e_{l_j}}]_{1\leq j \leq k}\mathbf{Q}'_{k \otimes}$ has full rank $kJ$. %
According to Theorem \ref{theorem: circular_shift_solutions_to_circular_shift_based_solution} in the previous subsection, when $(\mathbf{K}_{d,e})$ equipped with the source encoding matrix $\mathbf{P}_{k \otimes}$ is a linear solution at rate $J/L$, then ($\mathbf{P}\mathbf{K}_{d, e}\mathbf{Q}$) is a linear solution (at rate $1$). %
As long as the Combination Network is concerned, we can also prove the following. %
\begin{theorem}
\label{theorem: circular_shift_solutions_to_circular_shift_based_solution_GG} 
On the $(n, k)$-Combination Network, the followings are equivalent:
\begin{enumerate}
\item \label{item:1)}
the code $(\mathbf{K}_{d,e})$, equipped with the source encoding matrix $\mathbf{P}_{k\otimes}$, is a linear solution at rate $J/L$;
\item \label{item:2)}
the code $(\mathbf{P}\mathbf{K}_{d,e}\mathbf{Q})$ is a linear solution (at rate $1$); 
\item \label{item:3)}
the code $(\mathbf{P}\mathbf{K}_{d,e}\mathbf{Q}')$ is a linear solution (at rate $1$).
\end{enumerate}
\begin{IEEEproof}
Please refer to Appendix-\ref{appendix: circular_shift_solutions_to_circular_shift_based_solution_GG}.
\end{IEEEproof}
\end{theorem}


For the $J$-dimensional circular-shift-based vector linear network code $(\mathbf{P}\mathbf{K}_{d, e}\mathbf{Q}')$ induced from ($\mathbf{K}_{d, e}$), the following lemma not only explicitly characterizes the decoding matrix at a receiver, which is the inverse matrix of $\mathbf{P}_{k \otimes}[\mathbf{F}_{e_{l_j}}]_{1\leq j \leq k}\mathbf{Q}'_{k \otimes}$, but also has been utilized in the proof that $3)$ implies $1)$ in Theorem \ref{theorem: circular_shift_solutions_to_circular_shift_based_solution_GG}. %
Reset $\mathbf{\Psi}(x)$ to be such a $k \times k$ matrix in which every entry is a polynomial over GF($p$) that $\mathbf{\Psi}(\mathbf{C}_L) = [\mathbf{F}_{e_{l_j}}]_{1\leq j \leq k}$. %

\begin{lemma}
\label{lemma: decoding_matrix_for_GAGT}
Assume $\mathrm{rank}(\mathbf{\Psi}(\beta^j)) = k$ for all $j \in \mathcal{J}$. %
Define the counterpart $k \times k$ matrix $\mathbf{\Psi}''(x)$ in which every entry is a polynomial over GF($p$)
\begin{equation}
\label{eqn: Psi_double_quotation_inverse_def}
\mathbf{\Psi}''(x) = \mathrm{det}(\mathbf{\Psi}(x)\tau(x))^{p^{m_L}-2} \mathrm{Adj}(\mathbf{\Psi}(x)).
\end{equation}
We have 
\begin{equation}
\label{eqn: decoding_matrix_for_GAG}
(\mathbf{G}_{k \otimes}\mathbf{\Psi}(\mathbf{C}_L)\tau(\mathbf{C}_L)_{k \otimes}\mathbf{G}_{k \otimes}^\mathrm{T}) 
(\bar{\mathbf{H}}_{k \otimes}^\mathrm{T}\mathbf{\Psi}''(\mathbf{C}_L)\mathbf{H}_{k \otimes})= \mathbf{I}_{k J}.
\end{equation}
\begin{IEEEproof}
Please refer to Appendix-\ref{appendix: proof_GAGT_decoding_matrix}.
\end{IEEEproof}
\end{lemma}

Consider a $J$-dimensional circular-shift-based vector linear solution $(\mathbf{P}\mathbf{K}_{d, e}\mathbf{Q})$ or $(\mathbf{P}\mathbf{K}_{d, e}\mathbf{Q}')$ on the $(n, k)$-Combination Network. %
As every $k$ out of $n$ layer-$3$ nodes correspond to a receiver, for all $1 \leq l_1 < \ldots < l_k \leq n$, the $k$ source data units can be recovered from $\mathbf{m}_{e_{l_1}}, \ldots, \mathbf{m}_{e_{l_k}}$. %
Consequently, the global encoding kernels for edges $\mathbf{m}_{e_{l_1}}, \ldots, \mathbf{m}_{e_{l_k}}$ directly induces a $J$-dimensional $(n, k)$ $p$-ary MDS array code, which will be mathematically defined in Definition \ref{def:MDS_array_codes}. %
Moreover, Lemma \ref{lemma: decoding_matrix_for_GAH} and Lemma \ref{lemma: decoding_matrix_for_GAGT} explicitly characterize the decoding matrix at
a receiver in the $(n, k)$-Combination Network, which can also be used to justify the MDS property of the new array codes introduced in Section \ref{sec: Application of the new vector LNC schemes to array codes}, as well as to characterize their decoding matrices. %
This will be discussed in detail in Section \ref{sec: Application of the new vector LNC schemes to array codes}. %

\section{Application of the new vector LNC schemes to array codes}
\label{sec: Application of the new vector LNC schemes to array codes}
In this section, we first follow the algebraic framework of vector LNC to present a unified description of array codes by using block generator matrices. %
Then, we introduce two new types of $(k+r, k)$ $p$-ary MDS array codes with $r \leq 3$. Compared with classical  EVENODD codes \cite{Blaum-Evenodd-ToC95}-\cite{Blaum-01-Generalized-EvenOdd-Cahpter} and row-diagonal parity (RDP) codes \cite{RDP_2004}\cite{Blaum-ISIT06-generalized-RDP}, 
the largest possible $k$ supported by the new codes is significantly increased. %
Moreover, under the setting that  $p = 2$, $L$ is an odd prime, $\mathbf{G} = [\mathbf{I}_{L-1}~\mathbf{0}]$ and $\mathbf{H} = [\mathbf{I}_{L-1}~\mathbf{1}]^\mathrm{T}$, %
we present scheduling algorithms for the encoding process of the new codes, so that the new codes can asymptotically approach the optimal encoding complexity $2$ XORs per original data bit with increasing $k$ and $L$. %

\subsection{A unified description of array codes based on vector LNC}
\label{Sec:unified_description_based_on_vector_LNC}
Erasure-correction codes are widely adopted in distributed data storage systems to enhance data reliability, such as in Redundant Arrays of Inexpensive Disks (RAID) \cite{patterson1989-raid-introduction} and in Ceph \cite{weil2006ceph}\cite{lei2024fpga}. %
Array codes are a particular class of erasure-correction codes which only employs binary operations in the coding processes, so that it takes much lower computational complexity compared with conventional linear codes defined over an extension field. %
A $J$-dimensional $(n, k)$ systematic binary array code encodes $k$ original data units, represented as a $J$-dimensional binary sequence, into $n-k$ redundant data units of the same dimension. %
It satisfies the MDS property if the $k$ original data units can be recovered from any $k$ out of the $n$ data units. %
Although array codes are typically defined over GF($2$), their construction can be extended to over GF($p$) (See, e.g., \cite{x-tit}). %

To the best of our knowledge, most constructions of array codes, such as EVENODD codes \cite{Blaum-Evenodd-ToC95}-\cite{Blaum-01-Generalized-EvenOdd-Cahpter}, RDP codes \cite{RDP_2004}\cite{Blaum-ISIT06-generalized-RDP}, XI-Code \cite{Huang-TCom16-Xi-Code}, etc., %
were elaborated in an ad-hoc manner or based on linear codes over a specific polynomial ring. The design insight is not comprehensible for readers without much background on abstract algebra. %
Inspired by the algebraic framework of vector LNC, which subsumes array codes as special instances, we propose the following definition to characterize a $J$-dimensional $(n, k)$ $p$-ary array code.

\begin{definition}
\label{def:MDS_array_codes}
A $J$-dimensional $(n,k)$ $p$-ary array code is prescribed by a $k\times n$ block generator matrix $[\mathbf{K}_{i,j}]_{1\leq i \leq k, 1\leq j \leq n}$ in which every block entry $\mathbf{K}_{i,j}$, called \emph{encoding kernel}, is an $J\times J$ matrix over GF($p$). It is said to be MDS if for all possible $1 \leq l_1 < \ldots < l_k \leq n$, $\mathrm{rank}([\mathbf{K}_{i,l_j}]_{1 \leq i,j \leq k}) = kJ$.
\end{definition}

A key reason to describe $(n, k)$ $p$-ary array codes under the algebraic framework of vector LNC is to interpret different code structures in a unified and transparent way, so as to facilitate further design of  array codes with more flexible parameter settings. In particular, a number of well-known array codes can be illustrated under the framework of vector LNC. %

From now on, for a $J$-dimensional $(n, k)$ \emph{systematic} array code, we have $[\mathbf{K}_{i,j}]_{1 \leq i \leq k, 1 \leq j \leq k} = \mathbf{I}_{kJ}$, so the code structure will be characterized by $[\mathbf{K}_{i,j}]_{1 \leq i \leq k, k < j \leq n}$. %
In the following examples, unless stated otherwise, $L$ is assumed to be an arbitrary odd prime, $p=2$, $\mathbf{G} = [\mathbf{I}_{L-1}~\mathbf{0}]$, $\mathbf{H} = [\mathbf{I}_{L-1}~\mathbf{1}]^\mathrm{T}$, and all array codes are $J$-dimensional with $J = L - 1$.

\begin{example}
\label{example:Evenodd}
(\emph{EVENODD code and its generalization}) EVENODD code \cite{Blaum-Evenodd-ToC95} is a $(k+2, k)$ systematic MDS array code with $k \leq L$. By translating the original construction description, we can express its encoding kernels $\mathbf{K}_{i, k+j}$, $1 \leq i \leq k$, as
\begin{equation}
\label{eqn:EVENODD}
\mathbf{K}_{i,k+1} = \mathbf{I}_{L-1}, ~ \mathbf{K}_{i,k+2} = \mathbf{G}\mathbf{C}_L^{i-1}\mathbf{G}^\mathrm{T} + \mathbf{S}_{i-1},
\end{equation}
where $\mathbf{S}_{0} = \mathbf{0}$ and $\mathbf{S}_{i'}$, $1 \leq i' < k$, replaces the $(L - i')^{th}$ row in $\mathbf{S}_0$ by the all-one vector. %
For $k \leq L$ and $r \geq 2$, another class of classical $(k+r, k)$ systematic array codes was constructed in \cite{Blaum-TIT96-MDS-Generalized_EvenOdd}, which can be characterized as
\begin{equation}
\label{eqn:generalized_EVENODD}
\mathbf{K}_{i, k+j} = \mathbf{G}\mathbf{C}_L^{(i-1)(j-1)}\mathbf{H}, 1 \leq i \leq k, 1 \leq j \leq r.
\end{equation}
The systematic array codes defined in \eqref{eqn:generalized_EVENODD} are known as generalized EVENODD codes \cite{Blaum-01-Generalized-EvenOdd-Cahpter}, because they subsume the original $(k+2, k)$ EVENODD code as a special case, which can be readily observed by noticing
\[
\mathbf{G}\mathbf{C}_L^{i-1}\mathbf{G}^\mathrm{T} + \mathbf{S}_{i-1} = \mathbf{G}\mathbf{C}_L^{i-1}\mathbf{H}, \forall 1 \leq i \leq k.
\]
The generalized EVENODD codes are known to be MDS for all odd prime $L$ in the case of $r \leq 3$ and for some choices of $L$ in the case of $r \geq 4$ (See, e.g., \cite{Blaum-TIT96-MDS-Generalized_EvenOdd}\cite{Blaum-01-Generalized-EvenOdd-Cahpter}). \hfill $\blacksquare$
\end{example}

\begin{example}
\label{example:RDP}
(\emph{RDP code and its generalization}) RDP code \cite{RDP_2004} is a $(k+2, k)$ systematic MDS array code with $k < L$. We can express its encoding kernels $\mathbf{K}_{i, k+j}$, $1 \leq i \leq k$, as
\begin{equation}
\label{eqn:RDP}
\mathbf{K}_{i,k+1} = \mathbf{I}_{L-1}, ~ \mathbf{K}_{i,k+2} = \mathbf{G}(\mathbf{C}_L^{i-1}+\mathbf{C}_L^{L-1})\mathbf{G}^\mathrm{T}.
\end{equation}
RDP code was generalized in \cite{Blaum-ISIT06-generalized-RDP} to be applicable to more general $r \geq 2$. For $(k+r, k)$ generalized systematic RDP code with $k < L$ and $r \geq 2$, its encoding kernels $\mathbf{K}_{i,k+1} = \mathbf{I}_{L-1}$, and $\mathbf{K}_{i,k+j}$, $1 \leq i \leq k, 2 \leq j \leq r$, can be characterized as
\begin{equation}
\label{eqn:generalized_RDP}
\mathbf{K}_{i,k+j} = \mathbf{G}(\mathbf{C}_L^{(i-1)(j-1)}+\mathbf{C}_L^{(L-1)(j-1)})\mathbf{G}^\mathrm{T}.
\end{equation}
It was proved in \cite{Blaum-ISIT06-generalized-RDP} that for defined $k, r$ and $L$, if the $(k+r, k)$ generalized EVENODD code is MDS then so is the $(k+r, k)$ generalized RDP code. Thus, for $k < L$ and all odd prime $L$, the $(k+3, k)$ generalized RDP code is MDS. \hfill $\blacksquare$
\end{example}

\begin{example}
\label{example:XI-code}
(\emph{XI-Code}) XI-Code is an $(L+1, L-2)$ MDS array code for some choices of $L$. When $L = 5$, the block generator matrix $[\mathbf{K}_{ij}]_{1\leq i \leq 3, 1 \leq j \leq 6}$ of the $(6, 3)$ XI-Code \cite{Huang-TCom16-Xi-Code} is given by
\begin{equation*}
\label{eqn:XI-code}
\begin{split}
&[\mathbf{K}_{ij}]_{1\leq i \leq 3, 1 \leq j \leq 6} = \\
&\setlength\arraycolsep{2pt}\left[\begin{array}{c:c:c:c:c:c}
\mathbf{I}_4 & \begin{matrix} 1 & 0 & 0 & 0 \\ 0 & 0 & 0 & 0 \\ 0 & 0 & 0 & 0 \\ 0 & 0 & 0 & 1 \end{matrix}
& \begin{matrix} 0 & 0 & 0 & 0 \\ 1 & 0 & 0 & 0 \\ 0 & 0 & 0 & 1 \\ 0 & 0 & 0 & 0 \end{matrix}
& \begin{matrix} 0 & 0 & 0 & 0 \\ 0 & 0 & 0 & 1 \\ 1 & 0 & 0 & 0 \\ 0 & 0 & 0 & 0 \end{matrix}
& \begin{matrix} 0 & 0 & 0 & 1 \\ 0 & 0 & 0 & 0 \\ 0 & 0 & 0 & 0 \\ 1 & 0 & 0 & 0 \end{matrix}
& \mathbf{I}_4
\\ \hdashline
\mathbf{0} & \begin{matrix} 0 & 1 & 0 & 0 \\ 0 & 0 & 1 & 0 \\ 0 & 0 & 0 & 1 \\ 1 & 0 & 0 & 0 \end{matrix}
& \begin{matrix} 0 & 0 & 0 & 0 \\ 0 & 0 & 0 & 0 \\ 0 & 1 & 0 & 0 \\ 0 & 0 & 1 & 0 \end{matrix}
& \begin{matrix} 1 & 0 & 0 & 0 \\ 0 & 0 & 0 & 1 \\ 1 & 0 & 0 & 0 \\ 0 & 0 & 0 & 1 \end{matrix}
& \begin{matrix} 0 & 0 & 0 & 1 \\ 1 & 0 & 0 & 0 \\ 0 & 0 & 0 & 0 \\ 0 & 0 & 0 & 0 \end{matrix}
& \begin{matrix} 0 & 1 & 0 & 0 \\ 0 & 0 & 1 & 0 \\ 1 & 0 & 0 & 0 \\ 0 & 0 & 0 & 1 \end{matrix}
\\ \hdashline
\mathbf{0} & \begin{matrix} 0 & 0 & 0 & 0 \\ 0 & 0 & 0 & 0 \\ 1 & 0 & 0 & 0 \\ 0 & 0 & 0 & 1 \end{matrix}
& \begin{matrix} 0 & 0 & 0 & 1 \\ 1 & 0 & 0 & 0 \\ 0 & 0 & 0 & 1 \\ 1 & 0 & 0 & 0 \end{matrix}
& \begin{matrix} 0 & 1 & 0 & 0 \\ 0 & 0 & 1 & 0 \\ 0 & 0 & 0 & 0 \\ 0 & 0 & 0 & 0 \end{matrix}
& \begin{matrix} 1 & 0 & 0 & 0 \\ 0 & 0 & 0 & 1 \\ 0 & 1 & 0 & 0 \\ 0 & 0 & 1 & 0 \end{matrix}
& \begin{matrix} 1 & 0 & 0 & 0 \\ 0 & 0 & 0 & 1 \\ 0 & 1 & 0 & 0 \\ 0 & 0 & 1 & 0 \end{matrix}
\end{array}\right].
\end{split}
\end{equation*}
One can readily check that this code is MDS, while many of its nonzero encoding kernels are not full rank. XI-Code is not systematic and belongs to the type of lowest density array codes. It satisfies the weakly systematic property \cite{Cassuto09-lowest-density-MDS}, that is, when $[\mathbf{K}_{ij}]_{1\leq i \leq k, 1 \leq j \leq n}$ is regarded as a $k(L-1)\times (k+3)(L-1)$ binary matrix, the $k(L-1)\times k(L-1)$ identity matrix can be obtained from it by deleting $(n-k)(L-1)$ columns. \hfill $\blacksquare$
\end{example}

\begin{example}
\label{example:TIT}
Let $\mathcal{J}$ denote a subset of $\{0, 1, 2, \ldots, L-1\}$ that is closed under multiplication by $p$ modulo $L$, and $J = |\mathcal{J}|$. %
Consider the $J$-dimensional $(k+r, k)$ $p$-ary MDS array codes constructed in \cite{x-tit}. %
The encoding kernels $\mathbf{K}_{i, j}$, $1 \leq i \leq k$, $1 \leq j \leq k+r$ can be expressed as
\begin{equation}
\label{eqn:TIT}
\mathbf{K}_{i, j} = \mathbf{G}(f_i(\mathbf{C}_L)\tau(\mathbf{C}_L))^{j-1}\mathbf{G}^{\mathrm{T}},
\end{equation}
where $\mathbf{G} = [\mathbf{I}_J ~ \mathbf{0}]$, $\tau(x)$ is defined as \eqref{def: tau(x)},
and $k$ polynomials $f_1(x),\ldots, f_k(x)$ over GF($p$) of degree at most $L-1$ subject to $\gcd(f_i(x) - f_j(x), x^L-1) = 1$ for all $1 \leq i < j \leq k$. \hfill $\blacksquare$
\end{example}

\subsection{Construction of EVENODD-like and RDP-like codes}
\label{subsec:EVENODD-like RDP-like codes}
It is well known that the classical $(k+2, k)$ RDP code over GF($2$) can exactly achieve the optimal encoding complexity, and the classical $(k+2, k)$ EVENODD code over GF($2$)  can asymptotically approach the optimal encoding complexity with increasing $k$ and the code dimensional $L-1$. %
In spite of the low complexity advantage, a key limitation for these codes is that $L$ is an upper bound for the selection of $k$. %
As discussed in Section \ref{subsec: combination_network}, a $J$-dimensional circular-shift-based vector linear solution over GF$(p)$ for the $(n,k)$-Combination Network naturally induces a $J$-dimensional $(n, k)$ $p$-ary MDS array code. %
Consequently, in light of the $J$-dimensional circular-shift-based vector linear solutions $(\mathbf{P}\mathbf{K}_{d, e}\mathbf{Q})$ and  $(\mathbf{P}\mathbf{K}_{d, e}\mathbf{Q}')$ described in Section \ref{subsec: combination_network}, 
we shall introduce two new types of $(k+r, k)$ $p$-ary MDS array codes with $r \leq 3$. 

Let $\mathcal{J}=\{j: \gcd(j,L)=1 \}$ and $J=|\mathcal{J}|$. Assume $\mathbf{G}$ and $\mathbf{H}$ are constructed subject to \eqref{U_bar_J} and \eqref{eqn:H_H_bar_def}, and they are both matrices over GF($p$). %
Define  
\begin{equation}
\label{def:A1}
\hat{\mathcal{A}} = \{\sum\nolimits_{j = 0}^{m_L-1} a_j\mathbf{C}_L^j: a_j \in \mathrm{GF}(p) \}\backslash\{\mathbf{0}\},
\end{equation}
which contains  $p^{m_L} - 1$ matrices over GF($p$). %

\begin{definition}
\label{def:EVENODD_like_code_definition}
(\emph{EVENODD-like codes}) Let $k \leq p^{m_L} - 1$ and  $r \leq 3$. A $(k+r, k)$ \emph{EVENODD-like code} is a $p$-ary systematic array code with the encoding kernels $\mathbf{K}_{i, k+j}$ prescribed by
\begin{equation}
\label{eqn:EVENODD_like_code_definition}
\mathbf{K}_{i, k+j} = \mathbf{G}\mathbf{A}_i^{j-1}\mathbf{H},~1 \leq i \leq k,~1 \leq j \leq r,
\end{equation}
where $\mathbf{A}_1, \ldots, \mathbf{A}_k$ are $k$ distinct matrices in $\hat{\mathcal{A}}$.
\end{definition}

\begin{definition}
\label{def:RDP_like_code_definition}
(\emph{RDP-like codes}) Let $k \leq p^{m_L} - 1$ and  $r \leq 3$. A $(k+r, k)$ \emph{RDP-like code} is a $p$-ary systematic array code with the encoding kernels $\mathbf{K}_{i, k+j}$ prescribed by
\begin{equation}
\label{eqn:RDP_like_code_definition}
\mathbf{K}_{i, k+j} = \mathbf{G}(\mathbf{A}_i\tau(\mathbf{C}_L))^{j-1}\mathbf{G}^\mathrm{T},~1 \leq i \leq k,1 \leq j \leq r,
\end{equation}
where $\mathbf{A}_1, \ldots, \mathbf{A}_k$ are $k$ distinct matrices in $\hat{\mathcal{A}}$, and $\tau(x)$ is defined in \eqref{def: tau(x)}.
\end{definition}

The systematic array codes given in Definition \ref{def:EVENODD_like_code_definition} and in Definition \ref{def:RDP_like_code_definition} are respectively said to be EVENODD-like and RDP-like because their respective analogous structure to generalized EVENODD codes (compare \eqref{eqn:EVENODD_like_code_definition} with \eqref{eqn:generalized_EVENODD}) and to generalized RDP codes (compare \eqref{eqn:RDP_like_code_definition} with \eqref{eqn:generalized_RDP}). %
Same as in other systematic Vandermonde array codes, the two types of codes defined above follow the customary way to generate the first encoded redundant data unit, denoted by $\mathbf{p}$, as $\mathbf{p} = \sum\nolimits_{i = 1}^k \mathbf{m}_i$,
where $\mathbf{m}_1, \ldots, \mathbf{m}_k$ are $k$ original data units, each of which is a $J$-dimensional row vector over GF($p$). %

In the design of array codes, it is popular to choose encoding kernels of the form $\mathbf{G}f(\mathbf{C}_L)\mathbf{G}^\mathrm{T}$ for certain polynomials $f(x)$ over GF($p$) (See, e.g., the codes recently designed in \cite{x-tit}-\cite{zhe_TCOM24}). %
However, in these constructions, $\mathbf{G}$ is always set to $[\mathbf{I}_{L'}~\mathbf{0}]$ with $L'<L$. %
In comparison, in our formulation of RDP-like codes, the selection of $\mathbf{G}$ is more flexibility. %
For instance, in the present consideration, when $L = 7$ and $p = 2$, $\mathbf{G}$ can be set as both $[\mathbf{I}_6~\mathbf{0}]$ and $[\mathbf{1} ~ \mathbf{J}_6]$ with $\mathbf{J}_6$ representing the $6\times 6$ anti-diagonal identity matrix. 

Given $k$ polynomials $f_1(x), \ldots, f_k(x)$ over GF($p$), let $\mathbf{\Phi}(x)$ and $\mathbf{\Phi}'(x)$ respectively denote the following $k\times (k+3)$ matrices
\begin{equation}
\mathbf{\Phi}(x) = \begin{bmatrix}
1 & 0 & \ldots & 0 & 1 & f_1(x) & f_1(x)^2 \\
0 & 1 & \ddots & \vdots & 1 & f_2(x) & f_2(x)^2 \\
\vdots & \ddots & \ddots & 0 & \vdots & \vdots & \vdots \\
0 & \ldots & 0 & 1 & 1 & f_k(x) & f_k(x)^2 \end{bmatrix},
\end{equation}
\begin{equation}
\mathbf{\Phi}'(x) = \begin{bmatrix}
1 & 0 & \ldots & 0 & 1 & f_1(x)\tau(x) & (f_1(x)\tau(x))^2 \\
0 & 1 & \ddots & \vdots & 1 & f_2(x)\tau(x) & (f_2(x)\tau(x))^2 \\
\vdots & \ddots & \ddots & 0 & \vdots & \vdots & \vdots \\
0 & \ldots & 0 & 1 & 1 & f_k(x)\tau(x) & (f_k(x)\tau(x))^2 \end{bmatrix},
\end{equation}
where $\tau(x)$ is the polynomial over GF($p$) defined in \eqref{def: tau(x)}. %
Consequently, the block generator matrices for a $(k+3, k)$ EVENODD-like code and a $(k+3, k)$ RDP-like can be expressed as $\mathbf{G}_{k \otimes}\mathbf{\Phi}(\mathbf{C}_L)\mathbf{H}_{n \otimes}$ and $\mathbf{G}_{k \otimes}\mathbf{\Phi}'(\mathbf{C}_L)\mathbf{G}^\mathrm{T}_{n \otimes}$, respectively,  with $f_1(\mathbf{C}_L), \ldots, f_k(\mathbf{C}_L)\in \hat{\mathcal{A}}$. %
\begin{theorem}
\label{thm:EVENODD_RDP_MDS}
Every EVENODD-like code and every RDP-like code is MDS.
\begin{IEEEproof}
Set $k = p^{m_L}-1$. %
Consider the $L$-dimensional circular-shift linear network code $(\mathbf{K}_{d,e})$ with local encoding kernels selected from $\hat{\mathcal{A}}$. %
In light of the descriptions in Section \ref{subsec: combination_network}, every EVENODD-like code and every RDP-like code respectively corresponds to a $J$-dimensional circular-shift-based vector linear network code $(\mathbf{P}\mathbf{K}_{d, e}\mathbf{Q})$ and  $(\mathbf{P}\mathbf{K}_{d, e}\mathbf{Q}')$ induced from $(\mathbf{K}_{d,e})$, where the local encoding kernels are constructed according to \eqref{eqn:EVENODD_like_code_definition} and \eqref{eqn:RDP_like_code_definition}.
Consequently, in order to prove the MDS property of EVENODD-like codes and RDP-like codes, it is equivalent to prove that the corresponding codes $(\mathbf{P}\mathbf{K}_{d, e}\mathbf{Q})$ and $(\mathbf{P}\mathbf{K}_{d, e}\mathbf{Q}')$ are qualified as vector linear solutions (at rate $1$). %
According to Theorem \ref{theorem: circular_shift_solutions_to_circular_shift_based_solution_GG}, it suffices to prove that the code $(\mathbf{K}_{d,e})$ equipped with the source encoding matrix $\mathbf{G}_{k\otimes}$ is a linear solution at rate $J/L$, that is
\begin{equation}
\label{full rank}
\mathrm{rank}(\mathbf{G}_{k\otimes}\mathbf{\Phi}(\mathbf{C}_L)) = kJ. 
\end{equation}
Define the following set of $k$ polynomials over GF($p$)
\begin{equation*}
  \{f_{1}(x), \ldots, f_{k}(x)\} = \{\sum\nolimits_{j = 0}^{m_L-1} a_jx^j: a_j \in \mathrm{GF}(p) \}\backslash\{0\}. 
\end{equation*}
In the proof of Theorem \ref{theorem: circular_shift_solutions_to_circular_shift_based_solution}, it has been proved that \eqref{eqn: circular_shift_based_full_rank} holds if and only if \eqref{eqn:rank psi beta} holds. Consequently, in order to prove \eqref{full rank} holds, it remains to show that $\mathrm{rank}(\mathbf{\Phi}(\beta^j)) = k$ for all $j \in \mathcal{J}$. %
Consider an arbitrary $j \in \mathcal{J}$. It is well known that a sufficient condition for  $\mathrm{rank}(\mathbf{\Phi}(\beta^j)) = k$ is that $f_1(\beta^j), \ldots, f_k(\beta^j)$ are $k$ distinct nonzero elements in GF($p^{m_L}$) (See, e.g., \cite{Lin_TCom18}). When $p = 2$, $L$ is a prime and $\mathcal{J} = \{1, \ldots, L-1\}$, it has been proved in Theorem 6 of the earlier conference version \cite{sun2023unified} that $f_1(\beta^j), \ldots, f_k(\beta^j)$ are $k$ distinct nonzero elements in GF($2^{m_L}$). Similar property also holds for general prime $p$, odd integer $L$ and $\mathcal{J} =\{j: \gcd(j,L)=1\}$, whose proof is essentially same as the proof of Theorem 6 in \cite{sun2023unified} and thus omitted here. %
We have asserted that $\mathrm{rank}(\mathbf{\Phi}(\beta^j)) = k$ for all $j \in \mathcal{J}$. %
\end{IEEEproof}
\end{theorem}

Notice that the MDS property of EVENODD-like codes and RDP-like codes can also be proved from the perspective of linear codes over a polynomial ring, which again involves additional knowledge in abstract algebra. %
In comparison, the proof of Theorems \ref{thm:EVENODD_RDP_MDS} we present stems from the framework of circular-shift-based vector LNC and utilizes Theorem \ref{theorem: circular_shift_solutions_to_circular_shift_based_solution_GG} and the 
intrinsic rank equation \eqref{eqn:rank_equation}. So, it has its own merit and applies to general Vandermonde MDS array codes, including both EVENODD-like codes and RDP-like codes. %

The proposed EVENODD-like code and RDP-like code not only share the MDS property, but also support the largest possible $k$ among all $(k+3, k)$ systematic Vandermonde circulant MDS array codes, as explicated below. %
In comparison, the classical EVENODD code and RDP code require $k \leq L$.

\begin{definition}
\label{def:Vandermonde circulant MDS array_code}
A $J$-dimensional $(k+r, k)$ systematic $p$-ary array code is called a \emph{Vandermonde circulant array code} if its encoding kernels $\mathbf{K}_{i, k+j}$ can be expressed as
\begin{equation}
\label{eqn:Vandermonde_circulant_MDS_array_code}
\mathbf{K}_{i, k+j} = \mathbf{P}\mathbf{A}_i^{j-1}\mathbf{Q},~1 \leq i \leq k,~1 \leq j \leq r,
\end{equation}
where $\mathbf{A}_1, \ldots \mathbf{A}_k \in \{\sum\nolimits_{j=0}^{L-1}a_j\mathbf{C}_L^j: a_j \in \mathrm{GF}(p)\}$, and $\mathbf{P}$, $\mathbf{Q}$ are certain defined matrices over GF($p$) of respective size $J\times L$ and $L\times J$. %
A $J$-dimensional $(k+r, k)$ nonsystematic $p$-ary array code is called a \emph{Vandermonde circulant array code} if its encoding kernels $\mathbf{K}_{i, j}$ can be expressed as
\begin{equation}
\label{eqn:Vandermonde_circulant_MDS_array_code}
\mathbf{K}_{i, j} = \mathbf{P}\mathbf{A}_i^{j-1}\mathbf{Q},~1 \leq i \leq k,~1 \leq j \leq k+r,
\end{equation}
where $\mathbf{A}_1, \ldots \mathbf{A}_k \in \{\sum\nolimits_{j=0}^{L-1}a_j\mathbf{C}_L^j: a_j \in \mathrm{GF}(p)\}$, and $\mathbf{P}$, $\mathbf{Q}$ are certain defined matrices over GF($p$) of respective size $J\times L$ and $L\times J$.
\end{definition}

It is interesting to notice that every $(k+r, k)$ Vandermonde circulant array code induces a $J$-dimensional vector linear network code on the $(k+r,k)$-Combination Network, and such a network code is qualified to be a circular-shift-based vector linear network code. %
The next proposition shows that the largest possible $k$ supported by $J$-dimensional $(k+r, r)$ MDS Vandermonde circulant $p$-ary array codes is upper bounded by $p^{m_L}-1$.

\begin{proposition}
\label{prop: max_k}
Let $r \geq 2$. If a $J$-dimensional $(k+r, r)$ Vandermonde circulant $p$-ary array code, either systematic or nonsystematic, is MDS, then $k \leq p^{m_L}-1$.
\end{proposition}

Proposition \ref{prop: max_k} asserts that EVENODD-like codes and RDP-like codes constructed in this subsection attains the largest possible $k$ among all  Vandermonde circulant $p$-ary MDS array code. It is interesting to remark that the nonsystematic array code designed in \cite{x-tit} and reviewed in Example \ref{example:TIT} is also qualified as a Vandermonde circulant $p$-ary MDS array code, and it attains the largest possible $k$ too, whose optimality was not unveiled in \cite{x-tit}.

\subsection{Efficient encoding by scheduling}
In this subsection, assume $p = 2$, $\mathbf{G} = [\mathbf{I}_{J}~\mathbf{0}]$ and $\mathbf{H} = [\mathbf{I}_{J}~\mathbf{A}]^\mathrm{T}$, where $\mathbf{A}$ is a $J\times (L-J)$ binary matrix such that $\mathbf{G}$ and $\mathbf{H}$ satisfy the conditions \eqref{U_bar_J} and \eqref{eqn:H_H_bar_def}. %
A key advantage of systematic Vandermonde circulant MDS array codes is the highly efficient encoding process. For an $L$-dimensional row vector $\mathbf{m}$, we follow the conventional assumption that there is no computational complexity cost to compute $\mathbf{m}\mathbf{C}_L^j$ on $\mathbf{m}$. %
Motivated by the scheduling method proposed in \cite{Lin_TCom18} to encode $(k+3, k)$ triply extended Reed-Solomon codes, we next design a scheduling method to encode $J$-dimensional EVENODD-like codes and RDP-like codes. %

Write $m = \left\lfloor \log_2 k \right\rfloor$.  For $1 \leq i \leq k$, let $b_{i,m}\ldots b_{i,0}$ be the binary representation of $i$, that is, $\sum_{j=0}^{m} b_{i,j}2^j = i$. %
Consider the $(k+3, k)$ EVENODD-like MDS array code $\mathcal{C}_E$ prescribed by the binary matrices $\mathbf{A}_i = \sum\nolimits_{j = 0}^{m} b_{i,j}\mathbf{C}_L^j \in \hat{\mathcal{A}}$, $1 \leq i \leq k$, used in \eqref{eqn:EVENODD_like_code_definition}. %
Let $\mathbf{m}_1, \ldots, \mathbf{m}_k$ be $k$ original data units to be encoded, each of which is a $J$-dimensional binary row vector. Let $\mathbf{p}_E, \mathbf{q}_E, \mathbf{r}_E$ denote the $3$ encoded redundant data units of the EVENODD-like MDS array code, each of which is also a $J$-dimensional binary row vector. Thus,
\begin{equation}
\begin{split}
\label{eqn:coded_data_units_GAH}
\mathbf{p}_E = \sum_{i = 1}\nolimits^k \mathbf{m}_i, \mathbf{q}_E = \sum\nolimits_{i = 1}^k\sum_{j = 0}\nolimits^m b_{i,j}\mathbf{m}_i\mathbf{G}\mathbf{C}_L^j\mathbf{H}, \\
\mathbf{r}_E = \sum\nolimits_{i = 1}^k\sum_{j = 0}\nolimits^m b_{i,j}\mathbf{m}_i\mathbf{G}\mathbf{C}_L^{2j}\mathbf{H}.
\end{split}
\end{equation}
We next propose an algorithm to generate $\mathbf{p}_E, \mathbf{q}_E, \mathbf{r}_E$ by scheduling, so that the intermediate outcome in the process of generating $\mathbf{p}_E$ can be re-used in generating $\mathbf{q}_E$ and $\mathbf{r}_E$.

\begin{algorithm}
\label{alg:encoding_scheduling_GAH}
Given $k$ original data units, the next procedure generates $\mathbf{p}_E$, $\mathbf{q}_E$, $\mathbf{r}_E$ of the $(k+3,k)$ EVENODD-like code.
\begin{itemize}
\item As initialization, set $\mathbf{t}_{0,1}$ to be the $J$-dimensional binary row vector $\mathbf{0}$, and set $\mathbf{t}_{0,i+1} = \mathbf{m}_i$ for every $1 \leq i \leq k$. Moreover, set ${n_j} = \left\lceil \frac{k + 1}{2^j} \right\rceil$ for every $0 \leq j \leq m+1 = \lfloor \log_2 k \rfloor+1$.
\item For $1 \leq j \leq m+1$, iteratively set
\begin{equation}
\label{eqn:alg_compute_t}
\begin{split}
\mathbf{t}_{j,a} = &{\mathbf{t}_{j - 1,2a - 1}} + {\mathbf{t}_{j - 1,2a}},\forall 1 \leq a < n_j, \\
\mathbf{t}_{j,n_j} = &\left\{\begin{matrix}
\mathbf{t}_{j-1,2n_{j}-1} + \mathbf{t}_{j-1,2n_{j}}, & \mathrm{if}~n_{j-1}~\mathrm{is~even} \\
\mathbf{t}_{j-1,2n_{j}-1}, & \mathrm{otherwise}
\end{matrix}\right.
\end{split}
\end{equation}
$\mathbf{p}_E$ is then obtained by setting $\mathbf{p}_E = \mathbf{t}_{m+1, 1}$.
\item For $0 \leq j \leq m$, iteratively set
\begin{equation}
\label{eqn:alg_compute_s}
\mathbf{s}_j = \sum\nolimits_{a=1}^{\lfloor n_{j}/2 \rfloor} \mathbf{t}_{j,2a}.
\end{equation}
\item $\mathbf{q}_E, \mathbf{r}_E$ can be respectively obtained by setting
 \begin{equation}
 \label{eqn:alg_compute_qE}
    \mathbf{q}_E = (\sum_{j=0}^m \mathbf{s}_j\mathbf{G}\mathbf{C}_L^j)\mathbf{H}, \mathbf{r}_E =  (\sum_{j=0}^m \mathbf{s}_j\mathbf{G}\mathbf{C}_L^{2j})\mathbf{H}.
 \end{equation}
\end{itemize}
One may readily check that the generated $\mathbf{p}_E, \mathbf{q}_E, \mathbf{r}_E$ are same as the ones defined in \eqref{eqn:coded_data_units_GAH}. \hfill $\blacksquare$
\end{algorithm}
In Algorithm \ref{alg:encoding_scheduling_GAH}, it takes $(k-1)J$ XORs to compute \eqref{eqn:alg_compute_t}, in which we assume $\mathbf{t}_{1,1} = \mathbf{0} + \mathbf{t}_{0,2}$ involves no XOR. %
It takes $(k - 1 - m )J$ XORs to compute \eqref{eqn:alg_compute_s}, and additional $2(mL+h)$ XORs to obtain the desired $\mathbf{q}_E$, $\mathbf{r}_E$ from \eqref{eqn:alg_compute_qE}, where $h$ denotes the number of $1$'s in $\mathbf{A}$. %
Specifically, when $L$ is an odd prime, we have $J=L-1$, $\mathbf{H} = [\mathbf{I}_{L-1}~\mathbf{1}]^\mathrm{T}$ and $h=L-1$.

\begin{proposition}
Assume $L$ is an odd prime. For the $(L-1)$-dimensional $(k+r, k)$ EVENODD-like code with $r = 2, 3$, the encoding process in Algorithm \ref{alg:encoding_scheduling_GAH} takes $2-\frac{3-r}{k}+
  (\frac{r-2}{k} + \frac{r-1}{k(L-1)})\lfloor\log_2 k\rfloor$ XORs per original data bit.
\end{proposition}

Now consider the $(k+3, k)$ RDP-like code $\mathcal{C}_R$ prescribed by the binary matrices $\mathbf{A}_i=\delta_{i}\mathbf{I}_L + \sum\nolimits_{j = 0}^{m} b_{i,j}\mathbf{C}_L^{j+1}$, $1 \leq i \leq k$, used in \eqref{eqn:RDP_like_code_definition}, where $\delta_i$ is equal to $1$ if $\sum\nolimits_{j = 0}^{m} b_{i,j}$ is odd and equal to $0$ otherwise. Let $\mathbf{p}_R, \mathbf{q}_R, \mathbf{r}_R$ denote the $3$ encoded redundant data units of the RDP-like code. %
In order to compute $\mathbf{p}_R, \mathbf{q}_R, \mathbf{r}_R$, we can first adopt the same procedure as in Algorithm \ref{alg:encoding_scheduling_GAH} to obtain $\mathbf{t}_{j,a}$ in \eqref{eqn:alg_compute_t} and $\mathbf{s}_{j}$ \eqref{eqn:alg_compute_s}. We thus have $\mathbf{p}_R = \mathbf{p}_E = \mathbf{t}_{m+1, 1}$. %
Further set
\begin{equation}
\label{eqn:alg_compute_s_sum}
\mathbf{s} = \sum\nolimits_{j=0}^m \mathbf{s}_j.
\end{equation}
Then, $\mathbf{q}_R$, $\mathbf{r}_R$ can be respectively obtained by setting
\begin{equation}
\label{eqn:alg_compute_qR}
    \mathbf{q}_R = \mathbf{s} + \sum_{j=0}^m \mathbf{s}_j\mathbf{G}\mathbf{C}_L^{j+1}\mathbf{G}^\mathrm{T},
    \mathbf{r}_R = \mathbf{s} + \sum_{j=0}^m \mathbf{s}_j\mathbf{G}\mathbf{C}_L^{2(j+1)}\mathbf{G}^\mathrm{T}.
\end{equation}
It can be justified that $\mathbf{q}_R$ and $\mathbf{r}_R$ obtained in \eqref{eqn:alg_compute_qR} are exactly the desired encoded data units of the RDP-like code. %
In the above procedure, it takes $mJ$ XORs to obtain $\mathbf{s}$ in \eqref{eqn:alg_compute_s_sum}, and additional $2(m+1)J$ XORs to obtain the $\mathbf{q}_R$, $\mathbf{r}_R$ from \eqref{eqn:alg_compute_qR}.

\begin{proposition}
Assume $L$ is an odd prime. For the $(k+r, k)$ RDP-like code with $r = 2, 3$, $2-\frac{3-r}{k}+\frac{r-1}{k}\lfloor\log_2 k\rfloor$ XORs per original data bit are sufficient in the encoding process.
\end{proposition}

Table \ref{table:XOR_number_classical_array_codes} lists the number of XOR operations per original data bit required in encoding by different types of $(L-1)$-dimensional binary MDS array codes for the case that $L$ is an odd prime. To make the expression more concise, the listed XOR number, \emph{i.e.}, the total number of required XORs in encoding divided by $k(L-1)$, corresponds to the limiting case that $L-1$ tends to infinity. %
As summarized in Table \ref{table:XOR_number_classical_array_codes}, the encoding complexity of EVENODD-like codes and RDP-like codes are lower than classical EVENODD codes and RDP codes for the case $r = 3$. As proved in  \cite{Lin_TCom18}, the encoding process of any $(k+3, k)$ MDS code requires at least $2 - 2/k$ XORs per original data bit. It turns out that classical $(k+3, k)$ EVENODD codes and RDP codes, which respectively require $k \leq L$ and $k < L$, do not have optimal encoding complexity. In comparison, the EVENODD-like codes and RDP-like codes proposed in this paper not only supports larger $k$ as long as $k < 2^{m_L}$, but also asymptotically approach the optimal encoding complexity $2$ XORs per data bit with increasing $k$ and $L$. %

Notice that the scheduling algorithm adopted in \cite{x-tit} can also be applied to RDP-like codes. %
Although by utilizing both the scheduling algorithm in \cite{x-tit} and the one designed in this paper, RDP-like codes can asymptotically approach the optimal encoding complexity $2$ XORs per data bit with increasing $k$ and $L$, the implementation of these two algorithms is different. %
Specifically, for $1 \leq i \leq k$, $\mathbf{A}_i$ in the encoding kernels is constructed based on the binary representation of $i$ in this paper, while $\mathbf{A}_i$ is constructed by using a certain defined polynomial equation in \cite{x-tit}. %
In addition, when employing the scheduling algorithm designed in this paper, the encoding complexity of EVENODD-like codes is slightly lower than that of RDP-like codes.

\begin{table}[h]
\center
\caption{Constraints of $k$ and the number of XORs per original data bit required in encoding by different types of $(L-1)$-dimensional binary MDS array codes for the case that $L$ is an odd prime ($r = 2, 3$).}
\label{table:XOR_number_classical_array_codes}
\renewcommand{\arraystretch}{1.5}  
\begin{tabular}{|c|c|c|}
 \hline
  Type of array codes & Constraints on $k$ & \# XORs in encoding   \\
  \hline
  EVENODD \cite{Blaum-Evenodd-ToC95}-\cite{Blaum-01-Generalized-EvenOdd-Cahpter}  & $k \leq L$ & $r - \frac{1}{k}$  \\
  \hline
  RDP \cite{RDP_2004}\cite{Blaum-ISIT06-generalized-RDP} & $k < L$ & $r- \frac{r}{k}$   \\
  \hline
  EVENODD-like  & \multirow{2}*{$k \leq 2^{m_L} - 1$} & \multirow{2}*{$2-\frac{3-r}{k}+\frac{r-2}{k}\lfloor\log_2 k\rfloor$  } \\
  (Definition \ref{def:EVENODD_like_code_definition}) & & \\
  \hline
  RDP-like  & \multirow{2}*{$k \leq 2^{m_L} - 1$} & \multirow{2}*{$2-\frac{3-r}{k}+\frac{r-1}{k}\lfloor\log_2 k\rfloor$} \\
  (Definition \ref{def:RDP_like_code_definition})  & & \\
  \hline
\end{tabular}
\end{table}

\appendix
\subsection{Proof of Proposition \ref{prop: G H GF(p)}}
\label{appendix:proof_G_H_GF(p)}
Let $\mathbf{v}_j$ denote the $(j+1)^{st}$ column in the Vandermonde matrix $\mathbf{V}_L$, that is, $\mathbf{V}_L = [\mathbf{v}_j]_{0\leq j \leq L-1}$. %
First, assume $\mathbf{G}$ is over GF($p$) and $[\mathbf{u}_j']_{j \notin \cal J} = \mathbf{0}$. We will prove that $\mathbf{H}$ is over GF($p$). %
Observe
\begin{equation}
[\mathbf{u}_j]_{j\in {\cal J}} = L_{\mathrm{GF}(p)}^{-1}\mathbf{G}[\mathbf{v}_j]_{j\in {\cal J}}.
\end{equation}
Define a $J \times J$ matrix $\mathbf{A}$ in the following way
\begin{equation}
\label{eqn:A}
\begin{split}
\mathbf{A}=&[\mathbf{u}_j]_{j\in {\cal J}}[\mathbf{u}_j]_{j\in {\cal J}}^\mathrm{T}\\
=&L_{\mathrm{GF}(p)}^{-2}\mathbf{G}[\mathbf{v}_j]_{j\in {\cal J}}[\mathbf{v}_j]_{j\in {\cal J}}^\mathrm{T}\mathbf{G}^\mathrm{T}.
\end{split}
\end{equation}
It is not difficult to check that
\begin{equation}
\begin{split}
&[\mathbf{v}_j]_{j\in {\cal J}}[\mathbf{v}_j]_{j\in {\cal J}}^\mathrm{T}\\
=&[\beta^{mn}]_{0 \leq m\leq L-1, n \in \mathcal{J}}[\beta^{nm}]_{n \in \mathcal{J}, 0 \leq m\leq L-1}\\
=&[\sum\nolimits_{j\in \mathcal{J}}\beta^{(m+n)j}]_{0 \leq m, n \leq L-1}.
\end{split}
\end{equation}
Because
\[
(\sum\nolimits_{j\in \mathcal{J}}\beta^{(m+n)j})^p=\sum\nolimits_{j\in \mathcal{J}}\beta^{(m+n)pj}=\sum\nolimits_{j\in \mathcal{J}}\beta^{(m+n)j},
\]
where the first equality holds by \cite{Lidl_Niederreiter_1996} and the last equality holds as ${\cal J}$ is closed under multiplication by $p$ modulo $L$.
According to lemma 2.4 in \cite{berlekamp2015algebraic}, the entry $\sum\nolimits_{j\in \mathcal{J}}\beta^{(m+n)j}$ belongs to GF($p$).
Hence, the matrix $[\mathbf{v}_j]_{j\in {\cal J}}[\mathbf{v}_j]_{j\in {\cal J}}^\mathrm{T}$ is over GF($p$). %
As $\mathbf{G}$ is also over GF($p$), so is $\mathbf{A}$ based on \eqref{eqn:A}. %
By the definition of $\mathbf{U}$, $[\mathbf{u}_j]_{j\in {\cal J}}$ is full rank, and so is $\mathbf{A} = [\mathbf{u}_j]_{j\in {\cal J}}[\mathbf{u}_j]_{j\in {\cal J}}^\mathrm{T}$. %
We now have
\begin{equation}
\begin{split}
[\mathbf{u}'_j]_{j\in {\cal J}}^{\mathrm{T}} = &L_{\mathrm{GF}(p)}^{-1}[\mathbf{u}_j]_{j\in {\cal J}}^{-1}\\
=&L_{\mathrm{GF}(p)}^{-1}[\mathbf{u}_j]_{j\in {\cal J}}^\mathrm{T}\mathbf{A}^{-1}\\
=&L_{\mathrm{GF}(p)}^{-2}[\mathbf{v}_j]_{j\in {\cal J}}^\mathrm{T}\mathbf{G}^\mathrm{T}\mathbf{A}^{-1},
\end{split}
\end{equation}
where the first equality holds by the definition in \eqref{U'_def}. %
Since $[\mathbf{u}_j']_{j \notin \cal J} = \mathbf{0}$ by the assumption,
\begin{equation}
\begin{split}
\mathbf{H}=&\mathbf{V}_L\mathbf{U}'^{\mathrm{T}}\\
=&[\mathbf{v}_j]_{j\in {\cal J}}[\mathbf{u}'_j]_{j\in {\cal J}}^{\mathrm{T}}\\
=&L_{\mathrm{GF}(p)}^{-2}[\mathbf{v}_j]_{j\in {\cal J}}[\mathbf{v}_j]_{j\in {\cal J}}^\mathrm{T}\mathbf{G}^\mathrm{T}\mathbf{A}^{-1}.
\end{split}
\end{equation}
Since all of $[\mathbf{v}_j]_{j\in {\cal J}}[\mathbf{v}_j]_{j\in {\cal J}}^\mathrm{T}$, $\mathbf{G}$ and $\mathbf{A}$ are over GF($p$), $\mathbf{H}$ is over GF($p$) too.

Conversely, assume that $\mathbf{H} = \mathbf{V}_L\mathbf{U}'^{\mathrm{T}}$ is over GF($p$) and $[\mathbf{u}_j]_{j \notin \cal J} = \mathbf{0}$. %
Consider a new $J \times L$ matrix $\bar{\mathbf{G}} = \bar{\mathbf{U}}\mathbf{V}_L^{-1}$ with $\bar{\mathbf{U}} = \mathbf{U}'$, %
and the counterpart $L\times J$ matrix $\bar{\mathbf{H}} = \mathbf{V}_L\bar{\mathbf{U}}'^{\mathrm{T}}$ with $\bar{\mathbf{U}}' = [\bar{\mathbf{u}}_j']_{0\leq j \leq L-1}$ defined by $[\bar{\mathbf{u}}'_j]_{j\in \cal J}^{\mathrm{T}} = L_{\mathrm{GF}(p)}^{-1}[\bar{\mathbf{u}}_j]_{j\in {\cal J}}^{-1}$ and $[\bar{\mathbf{u}}_j']_{j \notin \cal J} = \mathbf{0}$. %
Notice that $[\bar{\mathbf{u}}'_j]_{j\in \cal J}^{\mathrm{T}} = L_{\mathrm{GF}(p)}^{-1}[\bar{\mathbf{u}}_j]_{j\in {\cal J}}^{-1} = L_{\mathrm{GF}(p)}^{-1}[\mathbf{u}_j']_{j\in {\cal J}}^{-1}  = [\mathbf{u}_j]_{j\in {\cal J}}^\mathrm{T}$, and hence $\bar{\mathbf{U}}' = \mathbf{U}$. Moreover, as $\mathbf{V}_L$ is symmetric,
\begin{equation}
\bar{\mathbf{G}} = L_{\mathrm{GF}(p)}^{-1}\mathbf{H}^{\mathrm{T}}\widetilde{\mathbf{V}}_L^{2}, \bar{\mathbf{H}} = L_{\mathrm{GF}(p)}^{-1}\mathbf{V}_L^2\mathbf{G}^{\mathrm{T}}.
\end{equation}
By \eqref{eqn:VL_tilde^2}, $\widetilde{\mathbf{V}}_L$ is over GF($p$). %
Since both $\mathbf{H}$ and $\widetilde{\mathbf{V}}_L^{2}$ are over GF($p$), so is $\bar{\mathbf{G}}$. According to what has been justified in the previous paragraph, the fact that $\bar{\mathbf{G}}$ is over GF($p$) and $[\bar{\mathbf{u}}_j']_{j \notin \cal J} = \mathbf{0}$ implies that $\bar{\mathbf{H}}$ is over GF($p$). We can now affirm that as desired, $\mathbf{G} = L_{\mathrm{GF}(p)}^{-1}\bar{\mathbf{H}}^\mathrm{T}\widetilde{\mathbf{V}}_L^{2}$ is over GF($p$).

\subsection{Proof of Proposition \ref{prop:GAH}}
\label{appendix:proof_GAH}
For a polynomial $f(x)$ over $\mathrm{GF}(p)$, let $f(\mathbf{\Lambda})$ denote the $L\times L$ diagonal matrix obtained by evaluation of $f(x)$ under the setting $x = \mathbf{\Lambda}$. Equivalently, the diagonal entries in the diagonal matrix $f(\mathbf{\Lambda})$ are equal to $f(1), f(\beta), \ldots, f(\beta^{L-1})$ respectively. %
Based on \eqref{eqn:GH_def}, we have
\begin{equation}
\mathbf{G}f(\mathbf{C}_L)\mathbf{H} = L_{\mathrm{GF}(p)}^{-1}(\mathbf{U}\widetilde{\mathbf{V}}_L)(\mathbf{V}_Lf(\mathbf{\Lambda})\widetilde{\mathbf{V}}_L)(\mathbf{V}_L\mathbf{U}'^{\mathrm{T}}),
\end{equation}
where $\mathbf{V}_L\widetilde{\mathbf{V}}_L = \widetilde{\mathbf{V}}_L\mathbf{V}_L = L_{\mathrm{GF}(p)}\mathbf{I}_L$, so that
\begin{equation}
\begin{split}
&(\mathbf{G}f_1(\mathbf{C}_L)\mathbf{H})(\mathbf{G}f_2(\mathbf{C}_L)\mathbf{H})\\
=&L_{\mathrm{GF}(p)}^{2}\mathbf{U}f_1(\mathbf{\Lambda})\mathbf{U}'^{\mathrm{T}}\mathbf{U}f_2(\mathbf{\Lambda})\mathbf{U}'^{\mathrm{T}}
\end{split}
\end{equation}
and
\begin{equation}
\mathbf{G}f_1(\mathbf{C}_L)f_2(\mathbf{C}_L)\mathbf{H}\\
=L_{\mathrm{GF}(p)}\mathbf{U}f_1(\mathbf{\Lambda})f_2(\mathbf{\Lambda})\mathbf{U}'^{\mathrm{T}}
\end{equation}
In order to prove \eqref{eqn:GAH}, it is equivalent to prove that
\begin{equation}
\label{eqn:GAH_prop_equivalent_to_prove}
L_{\mathrm{GF}(p)}\mathbf{U}f_1(\mathbf{\Lambda})\mathbf{U}'^{\mathrm{T}}\mathbf{U}f_2(\mathbf{\Lambda})\mathbf{U}'^{\mathrm{T}}=\mathbf{U}f_1(\mathbf{\Lambda})f_2(\mathbf{\Lambda})\mathbf{U}'^{\mathrm{T}}.
\end{equation} %
Notice that for any polynomial $f(x)$ over GF($p$),
\begin{equation}
\mathbf{U}f(\mathbf{\Lambda})\mathbf{U}'^{\mathrm{T}} = \sum\nolimits_{0\leq j \leq L-1}f(\beta^j)\mathbf{u}_j\mathbf{u}_j'^{\mathrm{T}}.
\end{equation}
Since $[\mathbf{u}_j]_{j\notin \cal J} = \mathbf{0}$ or $[\mathbf{u}_j']_{j \notin \cal J} = \mathbf{0}$ by the assumption in \eqref{j notin J}, we have
\begin{equation}
\label{eqn:UfU_expression}
\mathbf{U}f(\mathbf{\Lambda})\mathbf{U}'^{\mathrm{T}} = \sum\nolimits_{j \in \mathcal{J}}f(\beta^j)\mathbf{u}_j\mathbf{u}_j'^{\mathrm{T}}.
\end{equation}
Consequently,
\begin{equation}
\label{eqn:left}
\begin{split}
&\mathbf{U}f_1(\mathbf{\Lambda})\mathbf{U}'^{\mathrm{T}}\mathbf{U}f_2(\mathbf{\Lambda})\mathbf{U}'^{\mathrm{T}}\\
=&\left(\sum\nolimits_{j_1 \in \mathcal{J}}f_1(\beta^{j_1})\mathbf{u}_{j_1}\mathbf{u}_{j_1}'^{\mathrm{T}}\right)\left(\sum\nolimits_{j_2 \in \mathcal{J}}f_2(\beta^{j_2})\mathbf{u}_{j_2}\mathbf{u}_{j_2}'^{\mathrm{T}}\right).
\end{split}
\end{equation}
Since $[\mathbf{u}'_j]_{j\in \cal J}^{\mathrm{T}}[\mathbf{u}_j]_{j\in {\cal J}} = L_{\mathrm{GF}(p)}^{-1}\mathbf{I}_J$ by the condition in \eqref{U'_def}, for $j_1, j_2 \in \mathcal{J}$,
\begin{equation}
\mathbf{u}_{j_1}'^{\mathrm{T}}\mathbf{u}_{j_2}= \left\{\begin{matrix} L_{\mathrm{GF}(p)}^{-1}, ~ j_1 = j_2 \\ 0,~j_1 \neq j_2 \end{matrix} \right.
\end{equation}
Hence, \eqref{eqn:left} is reduced to
\begin{equation}
\label{eqn:right}
\begin{split}
&L_{\mathrm{GF}(p)}\mathbf{U}f_1(\mathbf{\Lambda})\mathbf{U}'^{\mathrm{T}}\mathbf{U}f_2(\mathbf{\Lambda})\mathbf{U}'^{\mathrm{T}}\\
= &\sum\nolimits_{j \in \mathcal{J}}f_1(\beta^{j})f_2(\beta^{j})\mathbf{u}_{j}\mathbf{u}_{j}'^{\mathrm{T}}.
\end{split}
\end{equation}
According to \eqref{eqn:UfU_expression}, the right-hand side of \eqref{eqn:right} is exactly equal to $\mathbf{U}f_1(\mathbf{\Lambda})f_2(\mathbf{\Lambda})\mathbf{U}'^{\mathrm{T}}$, that is, \eqref{eqn:GAH_prop_equivalent_to_prove} and thus \eqref{eqn:GAH} hold as desired.

We next prove \eqref{eqn:HAG} based on \eqref{eqn:GAH}. We use the same technique as that in proving Proposition \ref{prop: G H GF(p)} to define $\bar{\mathbf{G}} = \bar{\mathbf{U}}\widetilde{\mathbf{V}}_L$ with $\bar{\mathbf{U}} = \mathbf{U}'$, %
and $\bar{\mathbf{H}} = \mathbf{V}_L\bar{\mathbf{U}}'^{\mathrm{T}}$ with $\bar{\mathbf{U}}' = [\bar{\mathbf{u}}_j']_{0\leq j \leq L-1}$ defined by $[\bar{\mathbf{u}}'_j]_{j\in \cal J}^{\mathrm{T}} = [\bar{\mathbf{u}}_j]_{j\in {\cal J}}^{-1}$ and $[\bar{\mathbf{u}}_j']_{j \notin \cal J} = \mathbf{0}$. %
It turns out that
\begin{equation}
\label{eqn:G_bar_H_bar_definition}
\bar{\mathbf{G}} = L_{\mathrm{GF}(p)}^{-1}\mathbf{H}^{\mathrm{T}}\widetilde{\mathbf{V}}_L^{2}, \bar{\mathbf{H}} = L_{\mathrm{GF}(p)}^{-1}\mathbf{V}_L^2\mathbf{G}^{\mathrm{T}}.
\end{equation}
Notice that if we set $\mathbf{U} = \bar{\mathbf{U}}$, $\mathbf{U}' = \bar{\mathbf{U}}'$, $\mathbf{G} = \bar{\mathbf{G}}$, $\mathbf{H} = \bar{\mathbf{H}}$, then such defined $\mathbf{U}$, $\mathbf{U}'$, $\mathbf{G}$ and $\mathbf{H}$ satisfy \eqref{U'_def}-\eqref{eqn:GH_def}. %
According to \eqref{eqn:GAH}, which has been justified in the previous paragraph, we have
\begin{equation}
\label{eqn:G_barA_H_bar}
(\bar{\mathbf{G}}\bar{f}_1(\mathbf{C}_L)\bar{\mathbf{H}})(\bar{\mathbf{G}}\bar{f}_2(\mathbf{C}_L)\bar{\mathbf{H}}) = \bar{\mathbf{G}}\bar{f}_1(\mathbf{C}_L)\bar{f}_2(\mathbf{C}_L)\bar{\mathbf{H}},
\end{equation}
where $\bar{f}_1(x)$ and $\bar{f}_2(x)$ are arbitrary two polynomials over GF($p$). Now, for every polynomial $f(x)$ over GF($p$), let $\bar{f}(x) = x^Lf(x^{-1})$, so that
\begin{equation}
\begin{split}
\bar{f}(\mathbf{C}_L) = &L_{\mathrm{GF}(p)}^{-1}\mathbf{V}_L\bar{f}(\mathbf{\Lambda})\widetilde{\mathbf{V}}_L\\
= &L_{\mathrm{GF}(p)}^{-1}\mathbf{V}_Lf(\mathbf{\Lambda}')\widetilde{\mathbf{V}}_L,
\end{split}
\end{equation} %
where $f(\mathbf{\Lambda}')$ is the diagonal matrix with diagonal entries equal to $f(1), f(\beta^{L-1}), \ldots, f(\beta)$, respectively.
By \eqref{eqn:VL_tilde^2}, $\widetilde{\mathbf{V}}_L^{2} = \mathbf{V}_L^{2} = L_{\mathrm{GF}(p)}\begin{bmatrix}\begin{smallmatrix}
1&0 & \ldots & 0\\
0&0 & \ldots & 1\\
 \vdots & \vdots & \begin{sideways} $\ddots$ \end{sideways} &\vdots  \\
0&1&\ldots  &0
\end{smallmatrix}\end{bmatrix}$, and thus $L_{\mathrm{GF}(p)}^{-2}\widetilde{\mathbf{V}}_L^{2}f(\mathbf{\Lambda}')\mathbf{V}_L^{2} = f(\mathbf{\Lambda})$. Consequently,
\begin{equation}
\label{eqn:f_bar_CL_expression}
\begin{split}
\bar{f}(\mathbf{C}_L) = &L_{\mathrm{GF}(p)}^{-3}\mathbf{V}_L^{3}f(\mathbf{\Lambda})\widetilde{\mathbf{V}}_L^{3}\\
= &L_{\mathrm{GF}(p)}^{-2}\mathbf{V}_L^{2}f(\mathbf{C}_L)\widetilde{\mathbf{V}}_L^{2},
\end{split}
\end{equation}
\begin{equation}
\label{eqn:f1*f2}
\bar{f}_1(\mathbf{C}_L)\bar{f}_2(\mathbf{C}_L) = L_{\mathrm{GF}(p)}^{-2}\mathbf{V}_L^{2}f_1(\mathbf{C}_L)f_2(\mathbf{C}_L)\widetilde{\mathbf{V}}_L^{2}.
\end{equation}
Based on \eqref{eqn:G_bar_H_bar_definition}, \eqref{eqn:f_bar_CL_expression} and \eqref{eqn:f1*f2}, Eq. \eqref{eqn:G_barA_H_bar} can be expressed as
\begin{equation}
(\mathbf{H}^{\mathrm{T}}f_1(\mathbf{C}_L)\mathbf{G}^{\mathrm{T}})(\mathbf{H}^{\mathrm{T}}f_2(\mathbf{C}_L)\mathbf{G}^{\mathrm{T}}) = \mathbf{H}^{\mathrm{T}}f_1(\mathbf{C}_L)f_2(\mathbf{C}_L)\mathbf{G}^{\mathrm{T}},
\end{equation}
that is, Eq. \eqref{eqn:HAG} is proved.

\subsection{Proof of Corollary \ref{cor: GAH_I}}
\label{appendix:proof_GAH_I}
Based on \eqref{eqn:diagonalization}, \eqref{eqn:GH_def}, and \eqref{eqn:GAH}, the left-hand side of Eq. \eqref{eqn: GAHGDH = I} can be written as $L_{\mathrm{GF}(p)}\mathbf{U}f(\mathbf{\Lambda})d(\mathbf{\Lambda})\mathbf{U}'^{\mathrm{T}}$.
Since $[\mathbf{u}_j]_{j\notin \cal J} = \mathbf{0}$ or $[\mathbf{u}_j']_{j \notin \cal J} = \mathbf{0}$ by the assumption in \eqref{j notin J}, we have
\begin{equation}
\label{eqn: UfdU'_in_sum}
\mathbf{U}f(\mathbf{\Lambda})d(\mathbf{\Lambda})\mathbf{U}'^{\mathrm{T}} = \sum\nolimits_{j \in \mathcal{J}}f(\beta^j)d(\beta^j)\mathbf{u}_j\mathbf{u}_j'^{\mathrm{T}}.
\end{equation}
Since $f(\beta^j)$ and $d(\beta^j)$ are over GF($p^{m_L}$), and for $j \in \mathcal{J}, f(\beta^j) \neq 0$, $f(\beta^j)d(\beta^j) = f(\beta^j)^{p^{m_L}-1} = 1$. Therefore, Eq. \eqref{eqn: UfdU'_in_sum} can be expressed as
\begin{equation}
\sum\nolimits_{j \in \mathcal{J}}f(\beta^j)d(\beta^j)\mathbf{u}_j\mathbf{u}_j'^{\mathrm{T}} = [\mathbf{u}_j]_{j\in {\cal J}}[\mathbf{u}'_j]_{j\in \cal J}^{\mathrm{T}}.
\end{equation}
According to \eqref{U'_def} that $[\mathbf{u}'_j]_{j \in \cal J}^{\mathrm{T}} = L_{\mathrm{GF}(p)}^{-1}[\mathbf{u}_j]_{j \in {\cal J}}^{-1}$, we can now affirm that $[\mathbf{u}_j]_{j\in {\cal J}}[\mathbf{u}'_j]_{j\in \cal J}^{\mathrm{T}} = L_{\mathrm{GF}(p)}^{-1}\mathbf{I}_J$. Thus Eq. \eqref{eqn: GAHGDH = I} can be proved. \\

We next prove \eqref{eqn: H'AG'H'DG' = I}. Same as the method in the proof of \eqref{eqn: GAHGDH = I}, Eq. \eqref{eqn: H'AG'H'DG' = I} can be represented as $L_{\mathrm{GF}(p)}\sum\nolimits_{j \in \mathcal{J}}f(\beta^j)d(\beta^j)\mathbf{u}_j'\mathbf{u}_j^{\mathrm{T}}$, where $f(\beta^j)$ and $d(\beta^j)$ also satisfy $f(\beta^j)d(\beta^j) = f(\beta^j)^{p^{m_L}-1} = 1$ for $j \in {\cal J}$. Thus we have
\begin{equation}
\sum\nolimits_{j \in \mathcal{J}}f(\beta^j)d(\beta^j)\mathbf{u}_j'\mathbf{u}_j^{\mathrm{T}} = [\mathbf{u}'_j]_{j\in {\cal J}}[\mathbf{u}_j]_{j\in \cal J}^{\mathrm{T}}.
\end{equation}
Based on \eqref{U'_def} that $[\mathbf{u}'_j]_{j\in \cal J}^{\mathrm{T}} = L_{\mathrm{GF}(p)}^{-1}[\mathbf{u}_j]_{j \in {\cal J}}^{-1}$, we can now affirm that $[\mathbf{u}'_j]_{j \in {\cal J}}[\mathbf{u}_j]_{j \in \cal J}^{\mathrm{T}} = L_{\mathrm{GF}(p)}^{-1}\mathbf{I}_J$. Eq. \eqref{eqn: H'AG'H'DG' = I} is thus proved. 

\subsection{Proof of Theorem \ref{theorem: circular_shift_solutions_to_circular_shift_based_solution}}
\label{appendix: theorem: circular_shift_solutions_to_circular_shift_based_solution}
Assume that an $L$-dimensional circular-shift linear network code $(\mathbf{K}_{d,e})$ equipped with the source encoding matrix $\mathbf{P}_{\omega \otimes}$ is a linear solution at rate $J/L$. %
According to Lemma $3$ and Proposition $7$ in \cite{Tang_CL19_encoding_decoding_circular_shift_LNC}, for every receiver $t$, there is an $\omega \times \omega$ block matrix $\mathbf{D}_t$ over GF($p$) such that 
\begin{equation}
\label{eqn: circular_shift_decoding_matrix_GH}
\mathbf{P}_{\omega \otimes}[\mathbf{F}_e]_{e \in \mathrm{In}(t)}\mathbf{D}_t\mathbf{Q}_{\omega \otimes} = \mathbf{I}_{\omega J}.
\end{equation}
In order to prove that the induced code $(\mathbf{P}\mathbf{K}_{d,e}\mathbf{Q})$ is a linear solution, it is sufficient to prove that there is an $\omega \times \omega$ block matrix $\mathbf{D}'_t$ over GF($p$) such that 
\begin{equation*}
(\mathbf{P}_{\omega \otimes}[\mathbf{F}_e]_{e \in \mathrm{In}(t)}\mathbf{Q}_{\omega \otimes})(\mathbf{P}_{\omega \otimes}\mathbf{D}'_t\mathbf{Q}_{\omega \otimes}) = \mathbf{I}_{\omega J}.
\end{equation*}
Based on Proposition \ref{prop:GAH}, we can directly derive that 
\begin{equation*}
\begin{split}
&(\mathbf{P}_{\omega \otimes}[\mathbf{F}_e]_{e \in \mathrm{In}(t)}\mathbf{Q}_{\omega \otimes})(\mathbf{P}_{\omega \otimes}\mathbf{D}'_t\mathbf{Q}_{\omega \otimes}) \\
= &\mathbf{P}_{\omega \otimes}[\mathbf{F}_e]_{e \in \mathrm{In}(t)}\mathbf{D}'_t\mathbf{Q}_{\omega \otimes}.
\end{split}
\end{equation*}
Then based on \eqref{eqn: circular_shift_decoding_matrix_GH}, we can obtain $\mathbf{G}_{\omega \otimes}[\mathbf{F}_e]_{e \in \mathrm{In}(t)}\mathbf{D}'_t\mathbf{H}_{\omega \otimes} = \mathbf{I}_{\omega J}$
when $\mathbf{D}'_t = \mathbf{D}_t$.

Conversely, assume the $J$-dimensional circular-shift-based vector linear network code $(\mathbf{P}\mathbf{K}_{d,e}\mathbf{Q})$ is a linear solution, that is
\begin{equation}
\label{eqn: circular_shift_based_full_rank}
\mathrm{rank}(\mathbf{P}_{\omega \otimes}[\mathbf{F}_e]_{e \in \mathrm{In}(t)}\mathbf{Q}_{\omega \otimes}) = \omega J.
\end{equation}
Let $\mathbf{\Psi}(x)$ be such an $\omega \times \omega$ matrix in which every entry is a polynomial over GF($p$) such that $\mathbf{\Psi}(\mathbf{C}_L) = [\mathbf{F}_e]_{e \in \mathrm{In}(t)}$. %
Based on \eqref{eqn: circular_shift_based_full_rank} and the rank equation \eqref{eqn:rank_equation} in Section \ref{subsec: GH_linear_solution}, by adopting the similar approach in the proof of Theorem $4$ in \cite{Tang_Sun_Circular-shift_LNC_TCOM}, we can deduce that \eqref{eqn: circular_shift_based_full_rank} holds if and only if the following holds.  
\begin{equation}
\label{eqn:rank psi beta}
\mathrm{rank}(\mathbf{\Psi}(\beta^j)) = \omega,~\forall~j \in \mathcal{J}.
\end{equation} %
By further utilizing Lemma \ref{lemma: decoding_matrix_for_GAH} introduced in Section \ref{subsec: GH_linear_solution}, we assert that for every receiver $t$, there is an $\omega \times \omega$ matrix $\mathbf{D}'_t$ over GF($p$) such that 
\begin{equation*}
(\mathbf{P}_{\omega \otimes}[\mathbf{F}_e]_{e \in \mathrm{In}(t)}\mathbf{Q}_{\omega \otimes})(\mathbf{P}_{\omega \otimes}\mathbf{D}'_t\mathbf{Q}_{\omega \otimes}) = \mathbf{I}_{\omega J}.
\end{equation*}
Based on Proposition \ref{prop:GAH}, the following equation is established 
\begin{equation*}
\begin{split}
&(\mathbf{P}_{\omega \otimes}[\mathbf{F}_e]_{e \in \mathrm{In}(t)}\mathbf{Q}_{\omega \otimes})(\mathbf{P}_{\omega \otimes}\mathbf{D}'_t\mathbf{Q}_{\omega \otimes}) \\
= &\mathbf{P}_{\omega \otimes}[\mathbf{F}_e]_{e \in \mathrm{In}(t)}\mathbf{D}'_t\mathbf{Q}_{\omega \otimes} \\
= &\mathbf{I}_{\omega J}.
\end{split}
\end{equation*}
To prove that the $L$-dimensional circular-shift linear network code $(\mathbf{K}_{d,e})$ equipped with $\mathbf{P}_{\omega \otimes}$ is a linear solution at rate $J/L$, it is equivalent to prove that there is an $\omega \times \omega$ matrix $\mathbf{D}_t$ over GF($p$) such that
\begin{equation}
\label{eqn: circular_shift_based_to_circular_shift}
\mathbf{P}_{\omega \otimes}[\mathbf{F}_e]_{e \in \mathrm{In}(t)}\mathbf{D}_t\mathbf{Q}_{\omega \otimes} = \mathbf{I}_{\omega J}.
\end{equation}
Obviously, when $\mathbf{D}_t = \mathbf{D}'_t$, \eqref{eqn: circular_shift_based_to_circular_shift} holds.

\subsection{Proof of Lemma \ref{lemma: decoding_matrix_for_GAH}}
\label{appendix: proof_GAH_decoding_matrix}
Based on \eqref{eqn:GAH} and \eqref{eqn:HAG}, the left-hand side of Eq. \eqref{eqn: decoding_matrix_for_GAH} and Eq. \eqref{eqn: decoding_matrix_for_HTAGT} can be written as 
\begin{equation}
\label{eqn: Gotimes_A_Hotimes_reduced}
\begin{split}
&(\mathbf{G}_{\omega \otimes}\mathbf{\Psi}(\mathbf{C}_L)\mathbf{H}_{\omega \otimes}) (\mathbf{G}_{\omega \otimes}\mathbf{\Psi}'(\mathbf{C}_L)\mathbf{H}_{\omega \otimes}) \\  
=&\mathbf{G}_{\omega \otimes}\mathbf{\Psi}(\mathbf{C}_L)\mathbf{\Psi}'(\mathbf{C}_L)\mathbf{H}_{\omega \otimes}, 
\end{split}
\end{equation}
\begin{equation}
\label{eqn: HTotimes_A_GTotimes_reduced}
\begin{split}
&(\mathbf{H}_{\omega \otimes}^\mathrm{T}\mathbf{\Psi}(\mathbf{C}_L)\mathbf{G}_{\omega \otimes}^\mathrm{T}) (\mathbf{H}_{\omega \otimes}^\mathrm{T}\mathbf{\Psi}'(\mathbf{C}_L)\mathbf{G}_{\omega \otimes}^\mathrm{T}) \\
=&(\mathbf{H}_{\omega \otimes}^\mathrm{T}\mathbf{\Psi}(\mathbf{C}_L)\mathbf{\Psi}'(\mathbf{C}_L)\mathbf{G}_{\omega \otimes}^\mathrm{T}).
\end{split}
\end{equation}
Observe that 
\begin{equation}
\label{eqn: matrix_times_Adj_I}
\begin{split}
\mathbf{\Psi}(x)\mathbf{\Psi}'(x) = &\mathrm{det}(\mathbf{\Psi}(x))^{p^{m_L}-2} \mathbf{\Psi}(x)\mathrm{Adj}(\mathbf{\Psi}(x)) \\
= &\mathrm{det}(\mathbf{\Psi}(x))^{p^{m_L}-1}\mathbf{I}_{w},
\end{split}
\end{equation}
where the first equality holds by \eqref{eqn: Psi_inverse_def}, and the second equality holds by $\mathbf{\Psi}(x)\mathrm{Adj}(\mathbf{\Psi}(x)) = \mathrm{det}(\mathbf{\Psi}(x))\mathbf{I}_{w}$. %
For brevity, let $l(x)$ denote $\mathrm{det}(\mathbf{\Psi}(x))$. %
Consequently, \eqref{eqn: Gotimes_A_Hotimes_reduced} and \eqref{eqn: HTotimes_A_GTotimes_reduced} can respectively be written as 
\begin{equation}
\label{eqn: Gotimes_l_Hotimes_GAH}
\begin{split}
&\mathbf{G}_{\omega \otimes}\mathbf{\Psi}(\mathbf{C}_L)\mathbf{\Psi}'(\mathbf{C}_L)\mathbf{H}_{\omega \otimes} \\
= &\mathbf{G}_{\omega \otimes}\begin{bmatrix}\begin{smallmatrix}
l(\mathbf{C}_L)^{p^{m_L}-1} & & & \\
& l(\mathbf{C}_L)^{p^{m_L}-1} & & \\
& & & \ddots & \\
& & & & l(\mathbf{C}_L)^{p^{m_L}-1}
\end{smallmatrix}\end{bmatrix}\mathbf{H}_{\omega \otimes}.
\end{split}
\end{equation}
\begin{equation}
\label{eqn: HTotimes_l_GTotimes_GAH}
\begin{split}
&\mathbf{H}_{\omega \otimes}^\mathrm{T}\mathbf{\Psi}(\mathbf{C}_L)\mathbf{\Psi}'(\mathbf{C}_L)\mathbf{G}_{\omega \otimes}^\mathrm{T} \\
= &\mathbf{H}_{\omega \otimes}^\mathrm{T}\begin{bmatrix}\begin{smallmatrix}
l(\mathbf{C}_L)^{p^{m_L}-1} & & & \\
& l(\mathbf{C}_L)^{p^{m_L}-1} & & \\
& & & \ddots & \\
& & & & l(\mathbf{C}_L)^{p^{m_L}-1}
\end{smallmatrix}\end{bmatrix}\mathbf{G}_{\omega \otimes}^\mathrm{T}.
\end{split}
\end{equation}
Based on Corollary \ref{cor: GAH_I}, we can directly obtain $\mathbf{G}l(\mathbf{C}_L)^{p^{m_L}-1}\mathbf{H} = \mathbf{I}_{J}$ and 
$\mathbf{H}^\mathbf{T}l(\mathbf{C}_L)^{p^{m_L}-1}\mathbf{G}^\mathrm{T} = \mathbf{I}_{J}$, 
which further implies $\mathbf{G}_{\omega \otimes}\mathbf{\Psi}(\mathbf{C}_L)\mathbf{\Psi}'(\mathbf{C}_L)\mathbf{H}_{\omega \otimes} = \mathbf{I}_{\omega J}$ and 
$\mathbf{H}_{\omega \otimes}^\mathrm{T}\mathbf{\Psi}(\mathbf{C}_L)\mathbf{\Psi}'(\mathbf{C}_L)\mathbf{G}_{\omega \otimes}^\mathrm{T} = \mathbf{I}_{\omega J}$. %
Therefore, \eqref{eqn: decoding_matrix_for_GAH} and \eqref{eqn: decoding_matrix_for_HTAGT} can be proved. %

\subsection{Proof of Proposition \ref{prop: GkG'H''k'H=I}}
\label{appendix:proof_GkG'H''k'H_I}
First observe
\begin{equation}
\begin{split}
\mathbf{G}^\mathrm{T}\bar{\mathbf{H}}{^\mathrm{T}}=&\widetilde{\mathbf{V}}_L\mathbf{U}^\mathrm{T}\mathbf{U}''\mathbf{V}_L\\
=&L_{\mathrm{GF}(p)}^{-1}\widetilde{\mathbf{V}}_L(\mathbf{I}_L+\mathbf{X})\mathbf{V}_L,
\end{split}
\end{equation}
where $\mathbf{X} = L_{\mathrm{GF}(p)}\mathbf{U}^\mathrm{T}\mathbf{U}''+\mathbf{I}_L$. %
Consequently,
\begin{equation}
\begin{split}
&\mathbf{G}f_1(\mathbf{C}_L)\tau(\mathbf{C}_L)\mathbf{G}^\mathrm{T}\bar{\mathbf{H}}{^\mathrm{T}}f_2(\mathbf{C}_L)\mathbf{H}\\
=&\mathbf{G}f_1(\mathbf{C}_L)\tau(\mathbf{C}_L)f_2(\mathbf{C}_L)\mathbf{H}+\\
~&L_{\mathrm{GF}(p)}^{-1}\mathbf{G}f_1(\mathbf{C}_L)\tau(\mathbf{C}_L)\widetilde{\mathbf{V}}_L\mathbf{X}\mathbf{V}_Lf_2(\mathbf{C}_L)\mathbf{H}. %
\end{split}
\end{equation}
In order to prove \eqref{eqn:GAG'}, it suffices to prove that $\tau(\mathbf{C}_L)\widetilde{\mathbf{V}}_L\mathbf{X}=\mathbf{0}$. %
Based on \eqref{eqn:diagonalization} in preliminaries, we have
\begin{equation}
\label{eqn: tau_vl_x}
\begin{split}
\tau(\mathbf{C}_L)\widetilde{\mathbf{V}}_L\mathbf{X} = &L_{\mathrm{GF}(p)}^{-1}(\mathbf{V}_L\tau(\mathbf{\Lambda})\widetilde{\mathbf{V}}_L)\widetilde{\mathbf{V}}_L\mathbf{X}\\
= &L_{\mathrm{GF}(p)}^{-1}\mathbf{V}_L\tau(\mathbf{\Lambda})\widetilde{\mathbf{V}}_L^{2}\mathbf{X}.
\end{split}
\end{equation}
By \eqref{eqn:VL_tilde^2} in preliminaries, $\widetilde{\mathbf{V}}_L^{2}=L_{\mathrm{GF}(p)}\begin{bmatrix}\begin{smallmatrix}
1&0 & \ldots & 0\\
0&0 & \ldots & 1\\
 \vdots & \vdots & \begin{sideways} $\ddots$ \end{sideways} &\vdots  \\
0&1&\ldots  &0
\end{smallmatrix}\end{bmatrix}$. We thus obtain
\begin{equation}
\label{eqn: tau_bar}
\tau(\mathbf{\Lambda})\widetilde{\mathbf{V}}_L^{2}\mathbf{X} = \bar{\mathbf{\Lambda}}\mathbf{X},
\end{equation}
where $\bar{\mathbf{\Lambda}} = \tau(\mathbf{\Lambda})\widetilde{\mathbf{V}}_L^{2} = L_{\mathrm{GF}(p)}\begin{bmatrix}\begin{smallmatrix}
\tau(1)& &  & \\
&&  &\tau(\beta)\\
  &  & \begin{sideways} $\ddots$ \end{sideways} &  \\
&\tau(\beta^{L-1})&  &
\end{smallmatrix}\end{bmatrix}$. %
By the definition of $\tau(x)$ in \eqref{def: tau(x)} , we have $\tau(\beta^j)=0$ for all $j \notin \mathcal{J}$, which equivalently says $\tau(\beta^{L-j})=0$ for all $j\notin \bar{\mathcal{ J}}$, that is, the $(j+1)^{st}$ column in $\bar{\mathbf{\Lambda}}$ is the all-zero vector. %
In addition, observe that for $j\in \bar{\mathcal{ J}}$, the $(j+1)^{st}$ row in $\mathbf{X} = L_{\mathrm{GF}(p)}\mathbf{U}^\mathrm{T}\mathbf{U}''+\mathbf{I}_L$ is the all-zero vector. We can now affirm that $\bar{\mathbf{\Lambda}}\mathbf{X} = \mathbf{0}$, which further implies $\tau(\mathbf{C}_L)\widetilde{\mathbf{V}}_L\mathbf{X} = \mathbf{0}$ based on \eqref{eqn: tau_bar} and \eqref{eqn: tau_vl_x}. Eq. \eqref{eqn:GAG'} is thus proved.

\subsection{Proof of Theorem \ref{theorem: circular_shift_solutions_to_circular_shift_based_solution_GG}}
\label{appendix: circular_shift_solutions_to_circular_shift_based_solution_GG}
Since Theorem \ref{theorem: circular_shift_solutions_to_circular_shift_based_solution} has proved that \ref{item:1)}) is equivalent to \ref{item:2)}), it remains to prove that \ref{item:1)}) is equivalent to \ref{item:3)}). %
The proof is essentially same as the proof of Theorem \ref{theorem: circular_shift_solutions_to_circular_shift_based_solution}, except that Proposition \ref{prop:GAH} and Lemma \ref{lemma: decoding_matrix_for_GAH} used in 
Theorem \ref{theorem: circular_shift_solutions_to_circular_shift_based_solution} is substituted by Proposition \ref{prop: GkG'H''k'H=I} and Lemma \ref{lemma: decoding_matrix_for_GAGT}. %

Assume that an $L$-dimensional circular-shift linear network code $(\mathbf{K}_{d,e})$ equipped with the source encoding matrix $\mathbf{P}_{k\otimes}$ is a linear solution at rate $J/L$. %
For every receiver $t$, there is a $k \times k$ block matrix $\mathbf{D}_t$ over GF($p$) such that 
\begin{equation}
\mathbf{P}_{k\otimes}[\mathbf{F}_{e_{l_j}}]_{1\leq j \leq k}\mathbf{D}_t\mathbf{Q}_{k\otimes} = \mathbf{I}_{kJ}.
\end{equation}
In order to prove that the induced code $(\mathbf{P}\mathbf{K}_{d,e}\mathbf{Q}')$ is a linear solution, 
it suffices to prove that there is a $k \times k$ block matrix $\mathbf{D}''_t$ over GF($p$) such that 
\begin{equation}
\label{eqn:theorem 10_step_1}
(\mathbf{P}_{k \otimes}[\mathbf{F}_{e_{l_j}}]_{1\leq j \leq k}\mathbf{Q}'_{k \otimes})(\bar{\mathbf{H}}_{k \otimes}^\mathrm{T}\mathbf{D}''_t\mathbf{Q}_{k \otimes}) = \mathbf{I}_{k J}.
\end{equation}
Based on Proposition \ref{prop: GkG'H''k'H=I} in Appendix, we have
\begin{align*}
&(\mathbf{P}_{k \otimes}[\mathbf{F}_{e_{l_j}}]_{1\leq j \leq k}\mathbf{Q}'_{k \otimes})(\bar{\mathbf{H}}_{k \otimes}^\mathrm{T}\mathbf{D}''_t\mathbf{Q}_{k \otimes})\\
=& \mathbf{P}_{k \otimes}[\mathbf{F}_{e_{l_j}}]_{1\leq j \leq k}\tau(\mathbf{C}_L)_{k \otimes}\mathbf{D}''_t\mathbf{Q}_{k \otimes}.
\end{align*}
Thus, \eqref{eqn:theorem 10_step_1} holds when $\mathbf{D}''_t = \tau(\mathbf{C}_L)_{k \otimes}^{p^{m_L}-2}\mathbf{D}_t$. %

Conversely, assume that the code $(\mathbf{P}\mathbf{K}_{d,e}\mathbf{Q}')$ is a linear solution, that is
\begin{equation}
\label{eqn: full_rank}
\mathrm{rank}(\mathbf{P}_{k \otimes}[\mathbf{F}_{e_{l_j}}]_{1\leq j \leq k}\mathbf{Q}'_{k\otimes}) = kJ.
\end{equation}
Let $\mathbf{\Psi}(x)$ to be such a $k \times k$ matrix in which every entry is a polynomial over GF($p$) that $\mathbf{\Psi}(\mathbf{C}_L) = [\mathbf{F}_{e_{l_j}}]_{1\leq j \leq k}$. %
Same as in the proof of Theorem \ref{theorem: circular_shift_solutions_to_circular_shift_based_solution}, we can utilize \eqref{eqn: full_rank} and \eqref{eqn:rank_equation} to obtain $\mathrm{rank}(\mathbf{\Psi}(\beta^j)) = k$ for all $j \in \mathcal{J}$. %
By further utilizing Lemma \ref{lemma: decoding_matrix_for_GAGT} in Section \ref{subsec: combination_network}, we assert that for every receiver, there is a $k \times k$ block matrix $\mathbf{D}''_t$ over GF($p$) such that
\begin{equation*}
(\mathbf{P}_{k \otimes}[\mathbf{F}_{e_{l_j}}]_{1\leq j \leq k}\mathbf{Q}'_{k \otimes})(\bar{\mathbf{H}}_{k \otimes}^\mathrm{T}\mathbf{D}''_t\mathbf{Q}_{k \otimes}) = \mathbf{I}_{k J}.
\end{equation*}
To prove that $(\mathbf{K}_{d,e})$ equipped with source encoding matrix $\mathbf{P}_{k\otimes}$ is a linear solution, it is equivalent to prove that there is a $k\times k$ matrix $\mathbf{D}_t$ over GF($p$) such that
\begin{equation}
\label{eqn:theorem 10_step_2}
\mathbf{P}_{k \otimes}[\mathbf{F}_{e_{l_j}}]_{1\leq j \leq k}\mathbf{D}_t\mathbf{Q}_{k \otimes} = \mathbf{I}_{k J}.
\end{equation}
Based on Proposition \ref{prop: GkG'H''k'H=I} in Appendix, \eqref{eqn:theorem 10_step_2} holds when $\mathbf{D}_t = \tau(\mathbf{C}_L)_{k\otimes}\mathbf{D}''_t$.

\subsection{Proof of Lemma \ref{lemma: decoding_matrix_for_GAGT}}
\label{appendix: proof_GAGT_decoding_matrix}
Analogous to the proof of Lemma \ref{lemma: decoding_matrix_for_GAH}, the first step is to simplify Eq. \eqref{eqn: decoding_matrix_for_GAG}. %
Based on \eqref{eqn:GAG'}, the left-hand side of Eq. \eqref{eqn: decoding_matrix_for_GAG} can be respectively written as 
\begin{equation}
\label{eqn: Gotimes_A_Gotimes_reduced}
\begin{split}
&(\mathbf{G}_{k \otimes}\mathbf{\Psi}(\mathbf{C}_L)\tau(\mathbf{C}_L)\mathbf{G}_{k \otimes}^\mathrm{T}) (\bar{\mathbf{H}}_{k \otimes}\mathbf{\Psi}''(\mathbf{C}_L)\mathbf{H}_{k \otimes}) \\  
=&\mathbf{G}_{k \otimes}\mathbf{\Psi}(\mathbf{C}_L)\tau(\mathbf{C}_L)\mathbf{\Psi}''(\mathbf{C}_L)\mathbf{H}_{k \otimes}. 
\end{split}
\end{equation}
Let $\mathbf{\Theta}(x)$ denote the $\mathbf{\Psi}(x)\tau(x)$. Observe that 
\begin{equation}
\label{eqn: matrix_times_Adj_I_Psi_double_quotation}
\begin{split}
\mathbf{\Theta}(x)\mathbf{\Psi}''(x) = &\mathrm{det}(\mathbf{\Theta}(x))^{p^{m_L}-2} \mathbf{\Theta}(x)\mathrm{Adj}(\mathbf{\Psi}(x)) \\
= &\mathrm{det}(\mathbf{\Theta}(x))^{p^{m_L}-1}\mathbf{I}_{k},
\end{split}
\end{equation}
where the first equality holds by \eqref{eqn: Psi_double_quotation_inverse_def}, and the second equality holds by $\mathbf{\Psi}(x)''\tau(x)\mathrm{Adj}(\mathbf{\Psi}(x)) = \mathrm{det}(\mathbf{\Psi}(x)\tau(x))\mathbf{I}_{w}$. %
For brevity, let $l'(x)$ denote $\mathrm{det}(\mathbf{\Theta}(x))$. %
Consequently, Eq. \eqref{eqn: Gotimes_A_Gotimes_reduced} can be written as %
\begin{equation}
\label{eqn: Gotimes_l_Hotimes_GAG}
\begin{split}
&\mathbf{G}_{k \otimes}\mathbf{\Theta}(\mathbf{C}_L)\mathbf{\Psi}''(\mathbf{C}_L)\mathbf{H}_{k \otimes} \\
= &\mathbf{G}_{k \otimes}\begin{bmatrix}\begin{smallmatrix}
l'(\mathbf{C}_L)^{p^{m_L}-1} & & & \\
& l'(\mathbf{C}_L)^{p^{m_L}-1} & & \\
& & & \ddots & \\
& & & & l'(\mathbf{C}_L)^{p^{m_L}-1}
\end{smallmatrix}\end{bmatrix}\mathbf{H}_{k \otimes}.
\end{split}
\end{equation}
In order to prove \eqref{eqn: decoding_matrix_for_GAG}, it suffices to prove $\mathbf{G}l'(\mathbf{C}_L)^{p^{m_L}-1}\mathbf{H} = \mathbf{I}_J$. %
Then the following corollary is necessary. %
\begin{corollary}
\label{cor: GAG'_I}
Consider a polynomial $f(x)$ over GF($p$) subject to $f(\beta^j) \neq 0$ for all $j \in {\cal J}$. Let $d(x) = (f(x)\tau(x))^{p^{m_L}-2}$, so that the following equations hold
\begin{equation}
\label{eqn: GAG'H'DH = I}
(\mathbf{G}f(\mathbf{C}_L)\tau(\mathbf{C}_L)\mathbf{G}^\mathrm{T})(\bar{\mathbf{H}}{^\mathrm{T}}d(\mathbf{C}_L)\mathbf{H}) = \mathbf{I}_J,
\end{equation}
\begin{IEEEproof}
Please refer to Appendix-\ref{appendix:proof_GAG'_I}.
\end{IEEEproof}
\end{corollary}
According to Corollary \ref{cor: GAG'_I}, we have $\mathbf{G}l'(\mathbf{C}_L)^{p^{m_L}-1}\mathbf{H} = \mathbf{I}_{J}$ 
which further implies $\mathbf{G}_{k \otimes}\mathbf{\Psi}(\mathbf{C}_L)\mathbf{\Psi}''(\mathbf{C}_L)\mathbf{H}_{k \otimes} = \mathbf{I}_{k J}$. %
Therefore, \eqref{eqn: decoding_matrix_for_GAG} can be proved. %

\subsection{Proof of Corollary \ref{cor: GAG'_I}}
\label{appendix:proof_GAG'_I}
Based on \eqref{eqn:diagonalization}, \eqref{eqn:GH_def}, and \eqref{eqn:GAG'}, the left-hand side of Eq. \eqref{eqn: GAG'H'DH = I} can be expressed as 
\begin{equation}
\label{eqn: GftauGT_HTdH_in_U}
\begin{split}
&(\mathbf{G}f(\mathbf{C}_L)\tau(\mathbf{C}_L)\mathbf{G}^\mathrm{T})(\bar{\mathbf{H}}{^\mathrm{T}}d(\mathbf{C}_L)\mathbf{H}) \\
= &L_{\mathrm{GF}(p)}\mathbf{U}f(\mathbf{\Lambda})\tau(\mathbf{\Lambda})d(\mathbf{\Lambda})\mathbf{U}'^{\mathrm{T}}. 
\end{split}
\end{equation}
By the assumption in \eqref{U' new_def}, every column vector $\mathbf{u}'_j$ in $\mathbf{U}'$ is nonzero for all $j \notin  {\cal J}$. We thus have
\begin{equation}
\label{eqn: UftaudU'_in_sum}
\mathbf{U}f(\mathbf{\Lambda})\tau(\mathbf{\Lambda})d(\mathbf{\Lambda})\mathbf{U}'^{\mathrm{T}} = \sum\nolimits_{j \in \mathcal{J}}f(\beta^j)\tau(\beta^j)d(\beta^j)\mathbf{u}_j\mathbf{u}_j'^{\mathrm{T}}.
\end{equation}
Since the elements $f(\beta^j)\tau(\beta^j)$ and $d(\beta^j)$ are in GF($p^{m_L}$), and $f(\beta^j)\tau(\beta^j) \neq 0$ for $j \in \mathcal{J}$ by the definition of $f(x)$ and $\tau(x)$, we have $f(\beta^j)\tau(\beta^j)d(\beta^j) = (f(\beta^j)\tau(\beta^j))^{p^{m_L}-1} = 1$. Therefore, the right-hand side of Eq. \eqref{eqn: UftaudU'_in_sum} can be expressed as
\begin{equation}
\label{eqn: GAGT_sum_in_columnU}
\sum\nolimits_{j \in \mathcal{J}}f(\beta^j)\tau(\beta^j)d(\beta^j)\mathbf{u}_j\mathbf{u}_j'^{\mathrm{T}} = [\mathbf{u}_j]_{j\in {\cal J}}[\mathbf{u}'_j]_{j\in \cal J}^{\mathrm{T}}.
\end{equation}
According to \eqref{U' new_def} that $[\mathbf{u}'_j]_{j\in \cal J}^{\mathrm{T}} = L_{\mathrm{GF}(p)}^{-1}[\mathbf{u}_j]_{j\in {\cal J}}^{-1}$, we can now affirm that $[\mathbf{u}_j]_{j\in {\cal J}}[\mathbf{u}'_j]_{j\in \cal J}^{\mathrm{T}} = L_{\mathrm{GF}(p)}^{-1}\mathbf{I}_J$. Consequently, \eqref{eqn: GAGT_sum_in_columnU}, \eqref{eqn: UftaudU'_in_sum} and \eqref{eqn: GftauGT_HTdH_in_U} imply Eq. \eqref{eqn: GAG'H'DH = I}.

\end{document}